%% file: main.tex
\documentclass[aps,twocolumn,superscriptaddress,floatfix,prb,10pt]{revtex4-2}
\usepackage{graphicx}
\usepackage{braket}
\usepackage{amsmath}
\usepackage{amssymb}
\usepackage{orcidlink}
\usepackage{hyperref}

\begin{document}

\title{Polarons with arbitrary nonlinear electron-phonon interaction}

\author{Stefano~Ragni\,\orcidlink{0009-0003-5603-2968}}
\affiliation{Faculty of Physics, Computational Materials Physics, University of Vienna, Kolingasse 14-16, Vienna A-1090, Austria}
\author{Tomislav~Mi\v{s}ki\'c\,\orcidlink{0009-0008-6382-5962}}
\affiliation{Department for Research of Materials under Extreme Conditions, Institute of Physics, 10000 Zagreb, Croatia}
\author{Thomas~Hahn\,\orcidlink{0000-0003-4325-7634}}
\affiliation{Center for Computational Quantum Physics, Flatiron Institute,
162 Fifth Avenue, New York, NY 10010, USA}
\author{Nikolay~Prokof’ev\,\orcidlink{0000-0002-5832-0889}}
\affiliation{Department of Physics, University of Massachusetts, Amherst, Massachusetts 01003, USA}
\author{Osor~S.~Bari\v{s}i\'c\,\orcidlink{0000-0002-6514-9004}}
\affiliation{Department for Research of Materials under Extreme Conditions, Institute of Physics, 10000 Zagreb, Croatia}
\author{Naoto~Nagaosa}
\affiliation{RIKEN Center for Emergent Matter Science (CEMS),
Wako, Saitama 351-0198, Japan}
\affiliation{Fundamental Quantum Science Program (FQSP), TRIP Headquarters, 
RIKEN, Wako 351-0198, Japan}
\author{Cesare~Franchini\,\orcidlink{0000-0002-7990-2984}}
\email{cesare.franchini@univie.ac.at}
\affiliation{Faculty of Physics, Computational Materials Physics, University of Vienna, Kolingasse 14-16, Vienna A-1090, Austria}
\affiliation{Dipartimento di Fisica e Astronomia, Universit\`a  di Bologna, 40127 Bologna, Italy}
\author{Andrey~S.~Mishchenko\,\orcidlink{0000-0002-7626-7567}}
\email{andrey.mishchenko6363@gmail.com}
\affiliation{Department for Research of Materials under Extreme Conditions, Institute of Physics, 10000 Zagreb, Croatia}
\affiliation{RIKEN Center for Emergent Matter Science (CEMS),
Wako, Saitama 351-0198, Japan}

\begin{abstract}
We develop an exact computational method based on numerical X-propagators for solving polaron models with arbitrary nonlinear couplings of local vibration modes to the electron density and magnitude of the hopping amplitude. Our approach covers various polaron models, some of which were impossible to treat by any existing approximation-free techniques. Moreover, it remains efficient in the most
relevant but computationally challenging regime of phonon frequencies much smaller than the electron bandwidth.
As a case study, we consider the double-well
type nonlinear model with quadratic ($g_2<0$) and quartic ($g_4>0$) interactions describing a broad class of technologically important materials, such as quantum paraelectric compounds and halide perovskites. 
We observe, depending on the model parameters, three qualitatively different regimes: (i) quantum interplay of quartic and quadratic interactions which suppresses effects of the quadratic coupling, (ii) intermediate-coupling regime with exponential $\propto \exp(\alpha g_2 \Omega^{-1/4})$ scaling of the quasiparticle weight and mass renormalization, and (iii) strong-coupling asymptotic behavior.         
\end{abstract}

\maketitle

\section{Introduction}

The vast majority of the polaron literature considers linear coupling between electrons and displacements, $\{ x_i \}$, of lattice ions \cite{Appel, AlexBook, Gunn_2008, franchini_polarons_2021}. However, it was recently realized that there exist many cases where lattice displacements in response to the electron density on site $i$, $V_{\rm ep}(x_i)$, and/or hopping between site $i$ and $j$, $t(x_i,x_j)$, are better described by nonlinear functions $V_{\rm ep}(x)$ and $t(x,x')$: halide perovskites \cite{Schilcher,Saidi2016,Zacharias2023}, doped manganites \cite{manganites,Hoesch2013}, quantum paraelectrics \cite{Bilz, Kumar, Stemmer_2018, Ranalli2024}, and common crystal structures consisting of intercalated sublattices of heavy and light atoms \cite{AdolphsPRB, Adolphs2}.
The superconducting pairing mechanism based on the two-phonon exchange process originally suggested in Ref.~\cite{Ngai} was revisited recently to explain $T_c$ properties in SrTiO$_3$ \cite{Feigel, Volkov2022}.
A generic non-linear coupling $V_{\rm ep}(x)$ 
involves quadratic $x^2$ and higher order in $x$ dependencies, leading, e.g., 
to the double-well potential for ions in quantum paraelectrics and ferroelectrics \cite{Bussmann-Holder, AdolphsPRB, Adolphs2, Spaldin, Ranalli, Verdi}. 
Finally, material physics applications may require accurate
treatment of generic nonlinear electron-phonon interaction (EPI)
when the function $V_{\rm ep}(x)$ is a numerical result originating from first-principles calculations \cite{Houtput2024, Bianco}.
 
Several studies devoted to the nonlinear electron-lattice interaction were either based on various approximations or
limited to the weak-coupling regime: they include
Refs.~\cite{AdolphsEPL, AdolphsPRB, Adolphs2} employing the Momentum Average (MA) approximation \cite{MA1, MAImprov, OCHol_Ber}, variational solutions \cite{KUKLOV}, 
quantum \cite{Paleari2021} and determinant \cite{Li2015, Li2015bis, Dee2020} Monte Carlo calculations restricted to finite clusters, and perturbation theory treatment \cite{RISEBOROUGH}. 
Recently, an extended Fröhlich polaron model incorporating quadratic EPI was studied in Refs.~\cite{Hout21} and \cite{Klimin24} using the variational Feynman’s path integral \cite{Fey55} and other analytic methods.

However, approximations can be detrimental already in the well-characterized case of linear interactions, where moderate-to-strong coupling results in the breakdown of the quasiparticle picture because of the Mott-Ioffe-Regel limit violation \cite{HolstMobil, FrohMobil}. 
Approximate treatments of nonlinear EPI are even more doubtful because they can, in contrast to the well-conditioned case of linear EPI, cause system instability even for single polarons \cite{stefano23}. 
Therefore, an exact method capable of treating polarons with arbitrary nonlinear EPI is highly desirable for achieving a precise theoretical understanding of the system and its technological applications. 
Significant progress towards this goal was made recently by developing the Diagrammatic Monte Carlo (DMC) method in momentum representation for quadratic coupling \cite{stefano23}, and by combining the path-integral DMC with the X-representation of vibrational modes \cite{stefano23,x-representation22, stefano23}, which established an approximation-free framework for dealing with linear and quadratic EPI. However, the key element in the last approach, namely, an exact analytic expression for phonon propagators in the X-representation, is not available for arbitrary EPI potentials. To proceed with a precise treatment of generic first-principles EPI coupling profiles, e.g. double-well potentials important for understanding anomalous properties of polarons in ferroelectric and quantum paraelectric materials~\cite{Kluibenschedl2025,joseph2025}, one needs to overcome this crucial shortcoming.    

In this work, we develop a Numerical X-propagators (NXP) method suitable for studying arbitrary nonlinear coupling to local lattice displacements. After we describe the computational technique for a general form of coupling, we apply it to the double-well potential, a system of broad relevance and interest, particularly for anharmonic materials that have been the focus of recent studies~\cite{Ranalli, Ranalli2024, Verdi, Spaldin, Shin2021, He2020,Zhou2019}.
We show how exact and approximate treatments deviate from each other in the adiabatic limit, where the phonon frequency is sufficiently small compared to the bare electronic hopping rate, which is the most relevant parameter regime for real materials.
 
The manuscript is organized as follows: Section \ref{MoMe} defines the model (Subsection \ref{MoMo}) and methods, with Subsection \ref{xrep} outlining the role of X-propagators in the technique and Subsection \ref{xprop} describing the novel tabulation methodology for arbitrary electron-lattice potentials. Section \ref{resu} contains results on the ground state properties of the double-well model for two different sets of parameters in Subsections \ref{fixed_t} and \ref{fixed_omega}, respectively.
Subsection \ref{spectral} contains an analysis of excited states and spectral properties. We present our conclusions and outlook 
in Section \ref{conclu}.

\section{Model and Method}
\label{MoMe}

The general Hamiltonian, which can be treated by the method developed in this paper, is 
\begin{equation}
\widehat{H} = \sum_{\langle ij \rangle} -t_{ij}(x_i,x_j) \, c_i^{\dagger} c_j + 
\sum_i \widehat{H}_{\rm e-p}^{\rm (i)}(x_i) \; ,
\label{h-gen}    
\end{equation}
where $c_i^{\dagger}$ ($c_i$) is the creation (annihilation) operator of an electron on site $i$ and $t_{ij}(x_i,x_j)$ is the hopping amplitude between sites $i$ and $j$, which depends on the local vibrational coordinates $x_i$ and $x_j$, respectively.
We assume that $t_{ij}(x_i,x_j)>0$ for any $(x_i,x_j)$, i.e. lattice vibrations do not change the hopping sign. The second term, 
$\widehat{H}_{\rm e-p}^{\rm (i)}(x_i)$, is the on-site Hamiltonian, which consists of the harmonic term describing atomic vibration with frequency $\Omega$ in the absence of the electron and an additional arbitrary potential $V_{\rm ep}(x_i)$ present only when the electron occupies the site $i$ 
\begin{equation}
\widehat{H}_{\rm e-p}^{\rm (i)}(x_i) = 
\Omega b_i^{\dagger}b_i  + 
n_i \sum_{n=1}^{\infty} g_n (b_i^{\dagger}+b_i)^n \; ,
\label{loc1}    
\end{equation}
where $b_i^{\dagger}$ ($b_i$) is the creation (annihilation) operator of a local vibration mode with frequency $\Omega$ on site $i$ (we count harmonic oscillator energies from $\Omega/2$) and expansion coefficients $g_n$ reproduce the potential $V_{\rm ep}(x_i)$.
This term effects atomic vibrations only when the electron occupation number on site $i$, $n_i = c_i^{\dagger} c_i$, is nonzero. 
To keep the connection with the standard form of the Hamiltonian (\ref{h-gen}) in terms of the physical displacement coordinate $x_i = (b_i^{\dagger}+b_i)/\sqrt{2\Omega}$ (the oscillator mass $m$ and Planck constant $\hbar$ are set to unity), the same Hamiltonian can be written as  
\begin{equation}
\widehat{H}_{\rm e-p}^{\rm (i)}(x_i) = 
-\frac{1}{2} \frac{\partial^2}{\partial x_i^2}
+ V(x_i)
\label{loc2}    
\end{equation}
with potential
\begin{equation}
V(x) = \frac{\Omega^2}{2} x^2 + n_i V_{\rm ep}(x)
\label{loc3}    
\end{equation}
being the sum of the harmonic and electron density dependent terms. Consistency with Eq.~{\ref{loc1}} implies  
\begin{equation}
V_{\rm ep}(x) = \sum_{n=1}^{\infty} g_n (2\Omega)^{n/2} x^n \; .
\label{loc4}    
\end{equation}

This Hamiltonian covers many existing polaron models. 
For example, a displacement-independent first term and a linear $V_{\rm ep}(x_i)$ 
are known as the Holstein model. On the other hand, a displacement dependent $t_{ij}(x_i,x_j)$ and $V_{\rm ep}=0$ lead to the Bari\v{s}i\'c-Labbe-Friedel-Su-Schrieffer-Heeger (BLF-SSH) model \cite{BLF1,BLF2,BLF3,SSH}. Both Holstein \cite{BoTruBa99,Goodvin11,DoubleCloud12} and BLF-SSH
\cite{SSH2010} systems are solvable numerically by several techniques, especially in one dimension and for large values of the $\Omega /t$ ratio in the antiadiabatic regime.

However, to treat linear plus quadratic dependence of $V(x_i)$, the adiabatic regime ($\Omega/t \ll 1$), higher-dimensional lattices,
and arbitrary dependence of the hopping magnitude $t_{ij}(x_i,x_j)$ on lattice vibrations, the only approximation-free approach is the X-representation DMC method of Refs.~\cite{x-representation22,stefano23}. 
In this approach the electron path-integral is ``dressed" by propagators of displaced  vibrational modes in imaginary time (X-propagators): 
\begin{equation}
U(x_i(\tau_1),x_i(\tau_2),\tau) = 
\bra{x_i(\tau_1)} 
e^{ - (\tau_2-\tau_1) \widehat{H}_{\rm e-p}^{\rm (i)}(x_i) }
\ket{x_i(\tau_2)} \, ,
\label{xprop_dep}    
\end{equation}
where $\tau=\tau_2-\tau_1$. Previous studies were limited
to the linear plus quadratic dependence of $V(x_i)$ because only
in this case there exists an analytic expression for $U$. 

In the present work, we develop an NXP method that can be used for exact treatment of any potential $V(x_i)$ in the Hamiltonian (\ref{h-gen}), which, thus, makes this Hamiltonian exactly solvable for any form of the electron-phonon interaction.    
In particular, we study the case of the double-well potential, which is considered a property governing physical processes in several technologically critical correlated materials \cite{Schilcher, Bilz, AdolphsPRB, Bussmann-Holder, Houtput2024}. 

\subsection{Model}
\label{MoMo}

The minimal EPI model featuring the local double-well potential can be formulated as Eq.~(\ref{loc1}) where the EPI term has nonzero coupling constants only for quadratic, $g_2<0$, and quartic, $g_4>0$, powers of $x$. Hopping amplitudes $t$ in the one-dimensional lattice are nonzero only for nearest neighbors and are independent of local vibrational coordinates $x$. We restrict our study to the one-dimensional case because of the weak dependence of the polaron properties on dimensionality for nonlinear coupling \cite{stefano23}. 

The properties of the model are defined by three independent dimensionless parameters (taking hopping $t$ as the unit of energy), see Eqs.~(\ref{h-gen}) and (\ref{loc1}): adiabatic ratio $\Omega /t$, $g_2/t$, and $g_4/t$. Since there is no particularly insightful way to define a single effective coupling, we chose to work with the bare coupling constant similarly to Ref.~\cite{AdolphsPRB}. 
Qualitative behavior for different values of $g_4$ was already studied in \cite{AdolphsPRB}, therefore, we focus on the $g_2$-dependence for different $\Omega /t$ ratios all the way to the adiabatic limit $\Omega /t \ll 1$.

\begin{figure}
    \centering
    \includegraphics[width=0.99\linewidth]{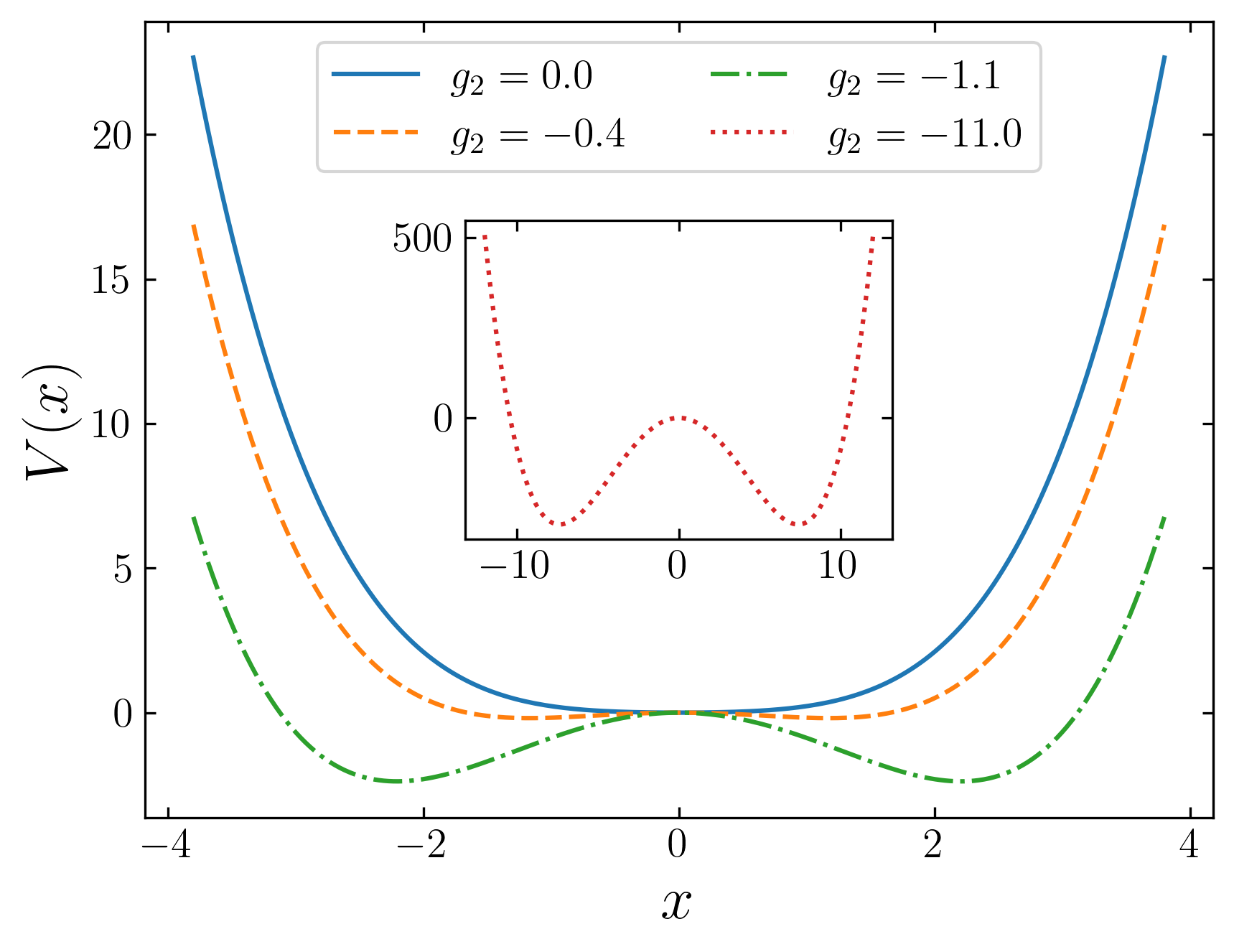}
    \caption{Potential energy profiles described by Eq.~\ref{loc3} when only $g_2$ and $g_4$ are nonzero in $V_{\rm ep}(x)$. $g_4 = 0.1$ and $\Omega = 0.5$ are kept fixed and four values of $g_2$ are shown, representative of different physical regimes that will be probed in the remainder of the text.}
    \label{fig:dw}
\end{figure}

A minimal graphical representation of the potential is shown in Fig.~\ref{fig:dw}. For $g_2 =0$ the potential is single-well.
When $g_2 < -\Omega/4$, the lattice becomes unstable. In this regime, the inclusion of anharmonic (quartic) terms stabilizes the system and gives rise to the characteristic double-well shape of the potential with two minima located at 
\begin{equation}
    x_0 = \pm \sqrt{\frac{-4 g_2-\Omega}{16 g_4 \Omega}}
\label{x00}    
\end{equation}
with the equilibrium potential energy 
\begin{equation}
E_{\pm} = - \frac{(4g_2/\Omega + 1)^2 \Omega^2}{64g_4} \; .
\label{epm}    
\end{equation}
The oscillator frequency at the minima is 
\begin{equation}
    \tilde\omega = R\Omega \; 
\label{tildeom}
\end{equation}
where
\begin{equation}
R = \sqrt{2(-4 g_2 / \Omega-1)} \; .
\label{RRR}
\end{equation}

As shown in Fig.~\ref{fig:dw}, decreasing $g_2$ leads to a higher energy barrier and a greater spatial separation between the two minima. The single-well ($g_2 = 0$) and shallow double-well ($g_2 = -0.4$) potentials are a defining feature of (quantum) paraelectrics. At the same time, deeper and more separated wells are typical of a ferroelectric material, such as $\text{BaTiO}_3$ below its Curie temperature \cite{Spaldin,joseph2025}. 

\subsection{X-representation}
\label{xrep}

The X-representation technique aims at computing the imaginary-time (Matsubara) Green's function \cite{x-representation22} 
\begin{equation}
    G_r(\tau > 0) = \langle c_{r}^{\,} (\tau) c_{0}^{\dagger} \rangle \, .
\end{equation}
The Hamiltonian in Eq.~\ref{h-gen} is split into the diagonal
(with respect to the electron occupation number) part
\begin{equation}
    \hat H_0 = \sum_i \widehat{H}_{\rm e-p}^{\rm (i)}(x_i) \, ,
\label{h00}
\end{equation}
and off-diagonal (hopping) part
\begin{equation}
    {\hat T} = \sum_{\langle ij \rangle}
    -t_{ij}(x_i,x_j) \, c_i^{\dagger} c_j \, .
\label{perrtt}
\end{equation}
Similarly to direct space DMC \cite{MacSawJar, Impurity}, one then carries out a perturbative expansion in powers of the electron hopping, considered formally as a Taylor series expansion of the so-called $\mathcal{T}$-exponent:
\begin{align}
    G_r(\tau > 0) = \langle c_{r}^{\,} (\tau) S(\tau) c_{0}^{\dagger} \rangle_0 \label{GFGF} \\
    S(\tau) = \mathcal{T} \exp{\left(-\int_0^\tau \text{d}\tau' \, {\hat T} (\tau')\right)}.
\end{align}

The zeroth order term yields the atomic limit (AL) Green's function
\begin{equation}\label{eq:gf_al}
    G_r^0 (\tau) = \delta_{r0}
    \langle e^{\tau \hat H_0} c_{r}^{\,} e^{-\tau \hat H_0} c_{0}^{\dagger} \rangle_0 \, .
\end{equation}
To represent the many-body state, we consider the following basis: $\ket{r,\{x_i\}}$, where $\ket{r}$ is an electronic orbital localized on site $r$ and $\ket{x_i}$ is an oscillator state on site $i$ with displacement $x_i$. Since we study a single electron in an otherwise empty band at temperature $T=1/\beta$, the average in Eq.~(\ref{eq:gf_al}) is done with respect to the thermodynamic state of all harmonic modes when $\{ n_i=0 \}$; we denote it as  $\ket{\varnothing}$. In this basis
\begin{gather}
    G_r^0 (\tau) = \int \text{d}x_0(0) \; \times \nonumber \\ \bra{\varnothing,x_0(0)} e^{-(\beta-\tau) \hat H_0} c_{r}^{\,} e^{-\tau \hat H_0} c_{0}^{\dagger} \ket{\varnothing,x_0(0)} \\ \label{eq:diagram}
    = \int \text{d}x_0(0) \int \text{d}x_0(\tau) \; 
    \times \nonumber 
    \\ U(x_0(0), x_0(\tau), \beta-\tau) \tilde U(x_0(\tau), x_0(0), \tau) \, , \label{eq:gf0_u}
\end{gather}
where we have inserted a complete set of vibrational basis states on site $0$, $\int \text{d}x_0(\tau) \ket{0,x_0(\tau)}\bra{0,x_0(\tau)}$.
The $U$ propagators on all empty sites can be ignored, because they reduce to partition function factors which cancel between the numerator and denominator in the average, or, equivalently, their integrated contribution is unity in the $T=0$ case considered in this work. Propagators $U$ and ${\tilde U}$ correspond to different local Hamiltonians, depending on the absence or presence of the electron on a given site, respectively.
Lastly, we include an electronic part containing an artificial ``chemical potential'' $\mu$,
\begin{equation}
G_\text{el}(\tau)=e^{\mu\tau} \, ,
\label{upgf}    
\end{equation}
to control the asymptotic decay of the solution in numerical simulations.

At first order, one hopping event occurs at time $\tau_1 < \tau $. Initially placed on site $i=0$, the electron hops to one of the nearest neighbor sites, say to site $i=1$. We repeat the procedure
of inserting a complete set of vibrational states on site $i=1$
to arrive at  
\begin{gather}
    \int_0^{\tau} \text{d}\tau_1
    \int \text{d}x_0(0) \int \text{d}x_0(\tau_1) \int \text{d}x_1(\tau_1) \int \text{d}x_1(\tau) \; 
    \times \nonumber \\
    U(x_1(\tau_1), x_1(\tau), \beta-\tau+\tau_1) \; \tilde U(x_1(\tau), x_1(\tau_1), \tau-\tau_1) \nonumber \\
    \tilde U(x_0(\tau_1), x_0(0), \tau_1) \;
    U(x_0(0), x_0(\tau_1), \beta-\tau_1) \;  \nonumber \\
    t_{10}(x_1(\tau_1),x_0(\tau_1))
    \times G_\text{el}(\tau) \, .
    \label{eq:gf1_u}
\end{gather}
In Eqs.~(\ref{eq:gf0_u}-\ref{eq:gf1_u}), we have cast the Green's function order-by-order contributions as products of elementary propagators, integrated over the basis states, which is reminiscent of the usual Feynman diagram representation. 

\begin{figure}
    \centering
    \includegraphics[width=0.99\linewidth]{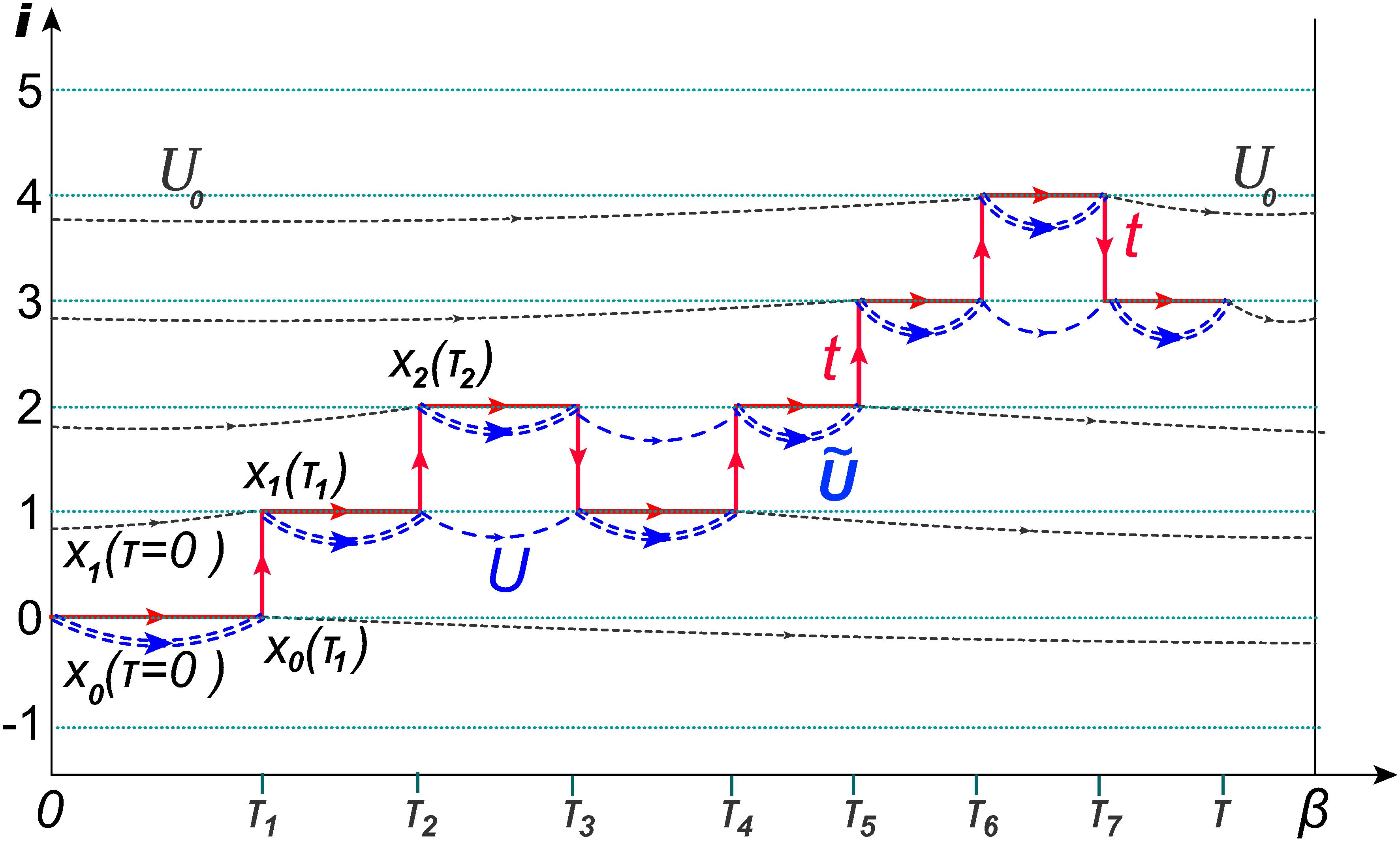}
    \caption{
    (modified from supplement in \cite{stefano23}). Green's function (\ref{GFGF}) diagrams in one-dimensional space (index $i$) and imaginary time ($\tau$) plane. Horizontal solid red lines with arrows are bare Green's functions (\ref{upgf}), while vertical lines with arrows are hopping events $t$. $\tilde{U}$ (double-line dashed arcs with arrows) and $U$ (single-line dashed arcs with arrows) are X-propagators with and without the electron on a given site, respectively. $U_0$ (dotted lines with arrows) is the limiting value of the X-propagator in the ground state technique (\ref{u0}).}
    \label{fig:xpropdia}
\end{figure}

A generic high-order expansion term is illustrated in Fig.~\ref{fig:xpropdia}. 
Detailed expressions and methods of calculating $U$ and $\tilde{U}$ propagators are discussed in subsection \ref{xprop} and Appendix \ref{xprop-app}. At $\beta \to \infty$, the X-propagators connected to the leftmost and rightmost parts of the electron path transform into their limiting ground state form $U_0$; see Appendix \ref{xprop-app}.
Each expansion term can be considered as a diagram with a well-defined mathematical weight given by a product of $G_\text{el}$, $t_{ij}$, $U_0$, $U$, and $\tilde U$ diagram elements. 
Diagrams can then be generated algorithmically by a Monte Carlo process using updates that add and remove hopping events and integrate over the displacement variables $x_0, x_1, \ldots, x_\tau$, and imaginary times $\tau_1, \tau_2, \ldots, \tau$ (see Fig.~\ref{fig:xpropdia}) \cite{stefano23}.

\subsection{X-propagators for an arbitrary interaction potential}
\label{xprop}

Previous applications of the X-representation technique \cite{stefano23,x-representation22} were limited to potentials with linear and quadratic terms that simply change the local Hamiltonian
from one harmonic form to another and, thus, admit an 
analytic solution for $\tilde{U}$, see Appendix \ref{analexp}. 
This limitation can be removed by precise tabulation of $\tilde{U}$ 
for an arbitrary potential. We propose two distinct methods for the numerical computation of $\tilde{U}$.

The first method is based on the Trotter-Suzuki decomposition \cite{trotter59, suzuki76}, to simulate the path-integral
of the vibrational coordinate \cite{Ceperley}
\begin{equation}
\begin{split}
& \;\;\;\;\;\;\;\; {\tilde U}(x_f,x_i,\tau)  = \int dx_1 dx_2 \ldots dx_{N-1} \times 
\nonumber \\ 
& \bra{x_f(\tau_f)}
e^{-\widehat{H}_{\rm e-p}^{\rm (i)}\delta\tau} 
\ket{x(\tau_{N-1})} \ldots 
\bra{x(\tau_1)} 
e^{-\widehat{H}_{\rm e-p}^{\rm (i)}\delta\tau} 
\ket{x(\tau_i)} \nonumber \, ,
\end{split}
\label{path1}    
\end{equation}
where the $\tau$ interval is divided into $N$ short
time slices $\delta\tau=\tau / N$ which allows decomposition of kinetic and potential energy and results in the following representation 
\begin{equation}
\begin{split}
& \;\;\;\;\;\;\;\; {\tilde U} (x_f,x_i,\tau)  = \int dx_1 dx_2 \ldots dx_{N-1} 
\times \nonumber \\ 
& P(x_f,x_{N-1},\delta\tau) \ldots
P(x_2,x_{1},\delta\tau) P(x_1,x_i,\delta\tau) 
\; ,
\label{path2} 
\end{split}
\end{equation}
where 
\begin{equation}
\begin{split}
& \;\;\;\;\;\;\;\;\; P(x,x',\delta\tau) = \frac{1}{\sqrt{2\pi\delta\tau}} \times  
\\
& \exp\left\{
-\delta\tau \left[
\frac{(x-x')^2}{2 \delta\tau^2}
 + \frac{V(x)+V(x')}{2} \right]
\right\} \; .
\end{split}
\label{path3}   
\end{equation}
is the free particle propagator in the presence of the (symmetrized) potential $[V(x)+V(x')]/2$.
One may significantly improve the efficiency of simulations by replacing 
$P(x,x',\delta\tau)$ with the exact harmonic oscillator solution (similar to $U(x,x',\delta\tau)$) for the potential
$V(\bar{x}) + V'(\bar{x}) (x-\bar{x}) + V''(\bar{x})(x-\bar{x})^2/2 $, where $\bar{x}=(x+x')/2$.

The Monte Carlo simulation of $\tilde{U}$ is then performed 
on the time mesh $\tau(i)= i \, \delta\tau$, up to some large 
imaginary time $\tau_{\rm  max}$, by sampling all possible values 
of $x$-coordinates on all time slices. The simplest set of updates for an arbitrary 
$V_{\rm ep}(x)$ is presented in Appendix \ref{TroSuz}.
The coordinates of the initial and final time slices are measured to collect statistics of $\tilde{U}$.
In practice, after introducing the coordinate 
mesh $x_j=dx(j-1/2)$ we collect statistics to bins of size $dx$. Let the corresponding histogram bin representing 
propagation from bin $j$ to bin $k$ in time $\tau_n$
be $S(j,k, n)$.  
The normalization is achieved by collecting statistics for $S(0,0,\delta \tau)$, which is also known extremely accurately 
analytically because at $x=0$ the quartic terms can be neglected. Thus, if 
\begin{equation}
{\cal M} = \int_{-dx/2}^{dx/2} dx \int_{-dx/2}^{dx/2} dx' 
P(x,x',\delta \tau )  \; ,
\label{norm1}    
\end{equation}
then, the properly normalized statistics for $\tilde{U}$
reads   
\begin{equation}
\tilde{U}(x_j,x_k,\tau_n ) = {\cal M} 
\frac{S(j,k,n)}{S(0,0,1)} \frac{1}{dx^2} \; .
\label{norm2}    
\end{equation}
Since we can afford to store rather dense meshes in memory, to  
incorporate numeric X-propagator to the DMC technique described in Subsection \ref{xrep}, we employ three 
linear space-time interpolations to get $\tilde{U}$ for 
$\tau > \delta \tau$, and 
\begin{equation}
{\cal U}(x,x',\tau)  = P(x,x',\tau)     
\end{equation}
for $\tau < \delta \tau $.

The second method to numerically tabulate the X-propagator is the single-site exact diagonalization (SSED) approach presented in Appendix \ref{ededed}. For the double-well potential studied in this work, it is more accurate and efficient than the general Monte Carlo approach. 
We benchmarked the two methods by comparison with analytic solutions for linear and quadratic cases, 
and cross-validated them by comparing results for the double-well potential. The only minor limitation of the exact diagonalization approach for a generic case is that it requires explicit expansion of the interaction potential into a series (\ref{loc4}) and may become less effective when the necessary expansion order is large. 
The SSED, used for calculating the X-propagator, mitigates the Hilbert space truncation that typically occurs when applying exact diagonalization to the entire lattice.
The remainder of the article uses the term ``exact diagonalization'' (ED) to refer to the approximate method of solving the problem of a polaron in the lattice with truncated Hilbert space. 

\begin{figure}
    \centering
    \includegraphics[width=0.99\linewidth]{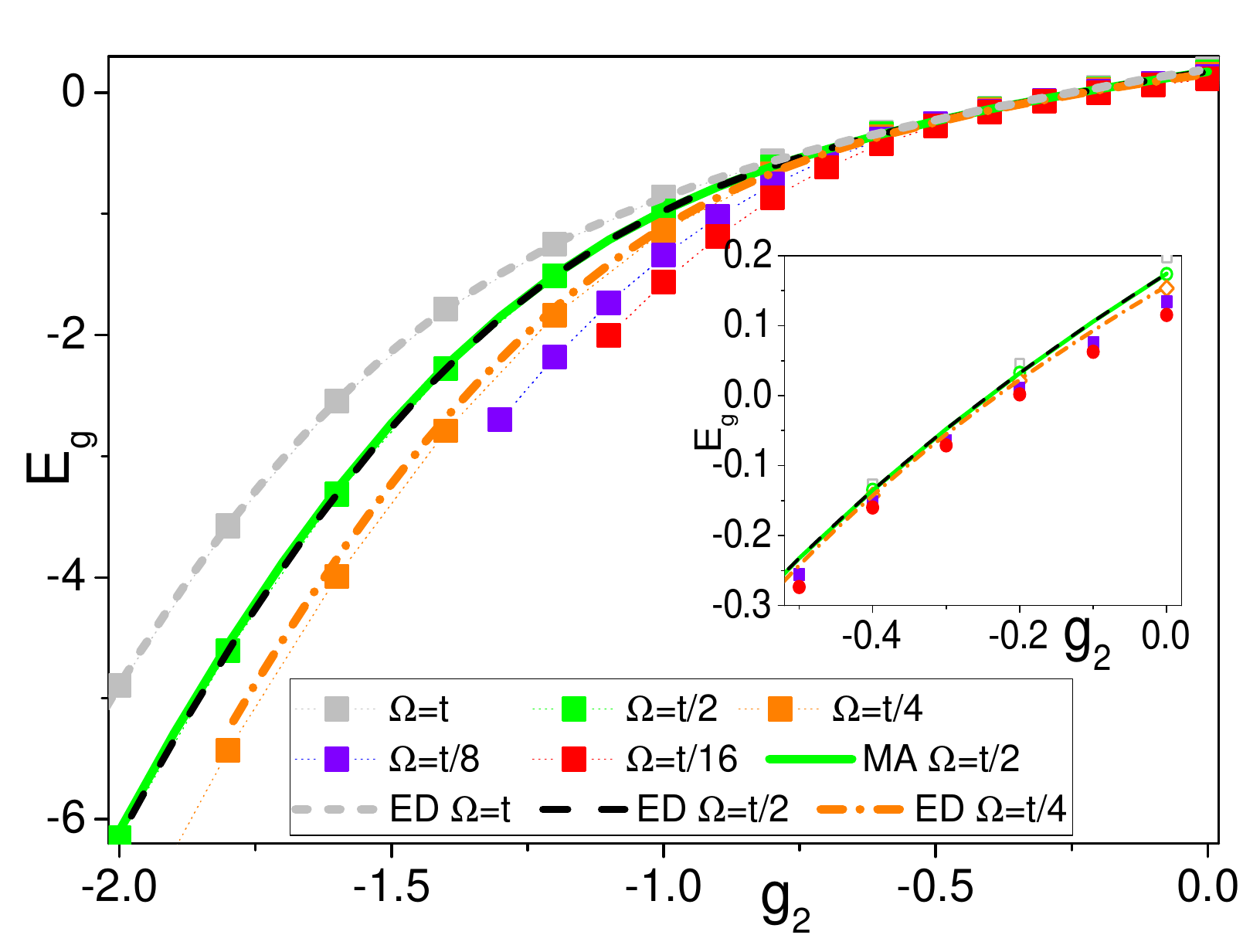}
    \caption{Energy renormalization for $t=1$, $g_4=0.1$ as a function of $g_2$ for different values of $\Omega$. Points are our numeric data. Error bars, if not shown, are smaller than the symbol size.
    Momentum Average (MA) method data \cite{AdolphsPRB} for $\Omega=t/2$ are shown by the green solid line, and Exact Diagonalization (ED) method data for $\Omega=t$, $\Omega=t/2$, and $\Omega=t/4$ are shown by the gray dotted, black dashed, and orange dash-dotted lines, respectively. }
    \label{fig:e}
\end{figure}

\section{Results}
\label{resu}

\begin{figure}
    \centering
    \includegraphics[width=0.99\linewidth]{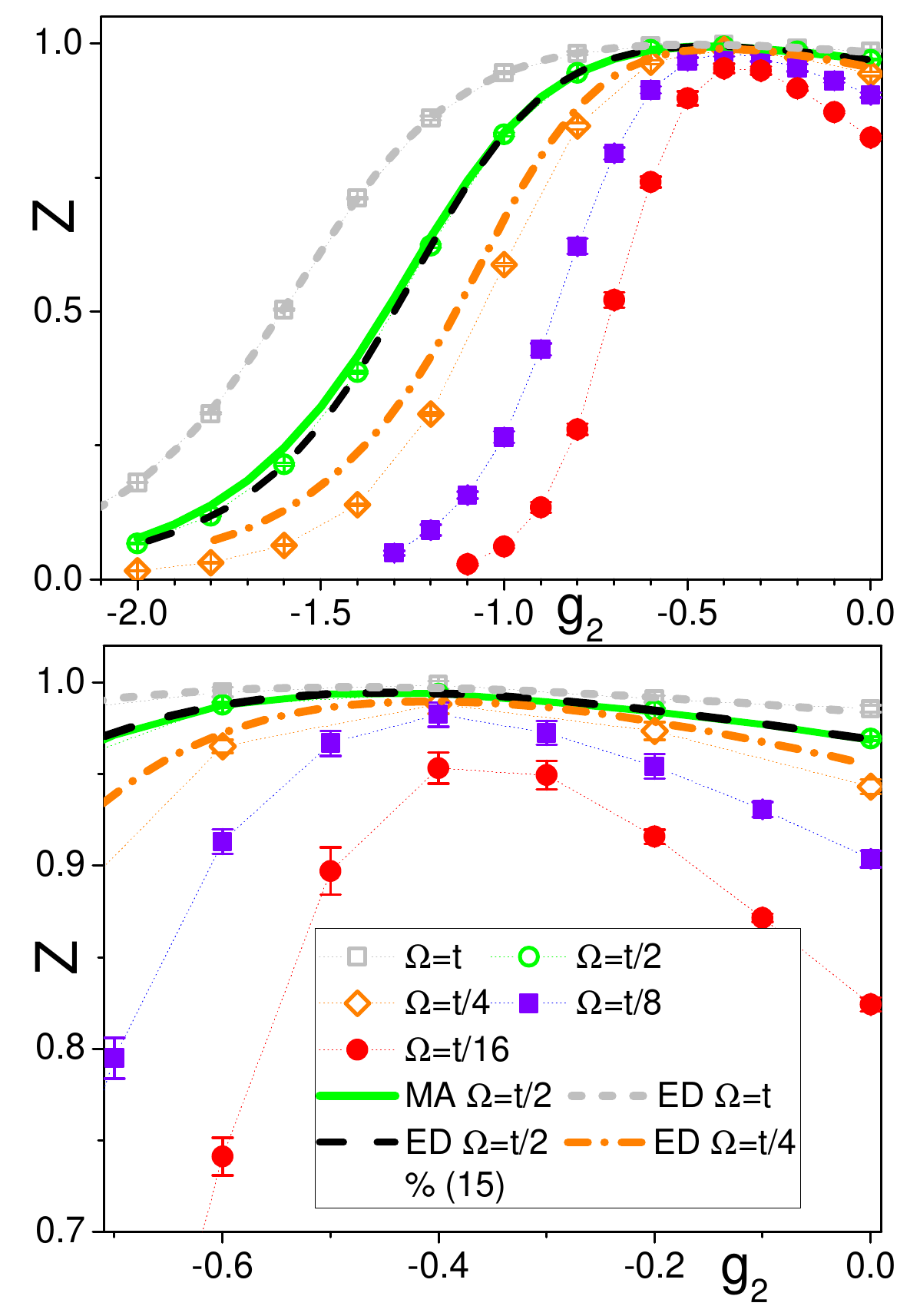}
    \caption{Quasiparticle weight $Z$. For notations, see the Fig.~\ref{fig:e} caption.
    Upper panel data are shown over a broad range of $g_2/t$ variation. Lower panel is focusing on the 
    small $g_2$ region.}
    \label{fig:z_be_ed}
\end{figure}

\subsection{Ground state properties at fixed \texorpdfstring{$t$}{t} and \texorpdfstring{$g_4$}{g4}}
\label{fixed_t}

This subsection presents results obtained by fixing the value of $g_4/t=0.1$, and exploring ground state properties as a function of $g_2/t$ for different values of $\Omega = t, t/2, t/4, t/8, t/16$. In Fig.~\ref{fig:e} we plot polaron energy counted from the bottom of the one-dimensional band at $-2t$ (when all interaction parameters are set to zero). The quasiparticle weight $Z$ is shown in Fig.~\ref{fig:z_be_ed}, and  
\begin{figure}
    \centering
    \includegraphics[width=0.99\linewidth]{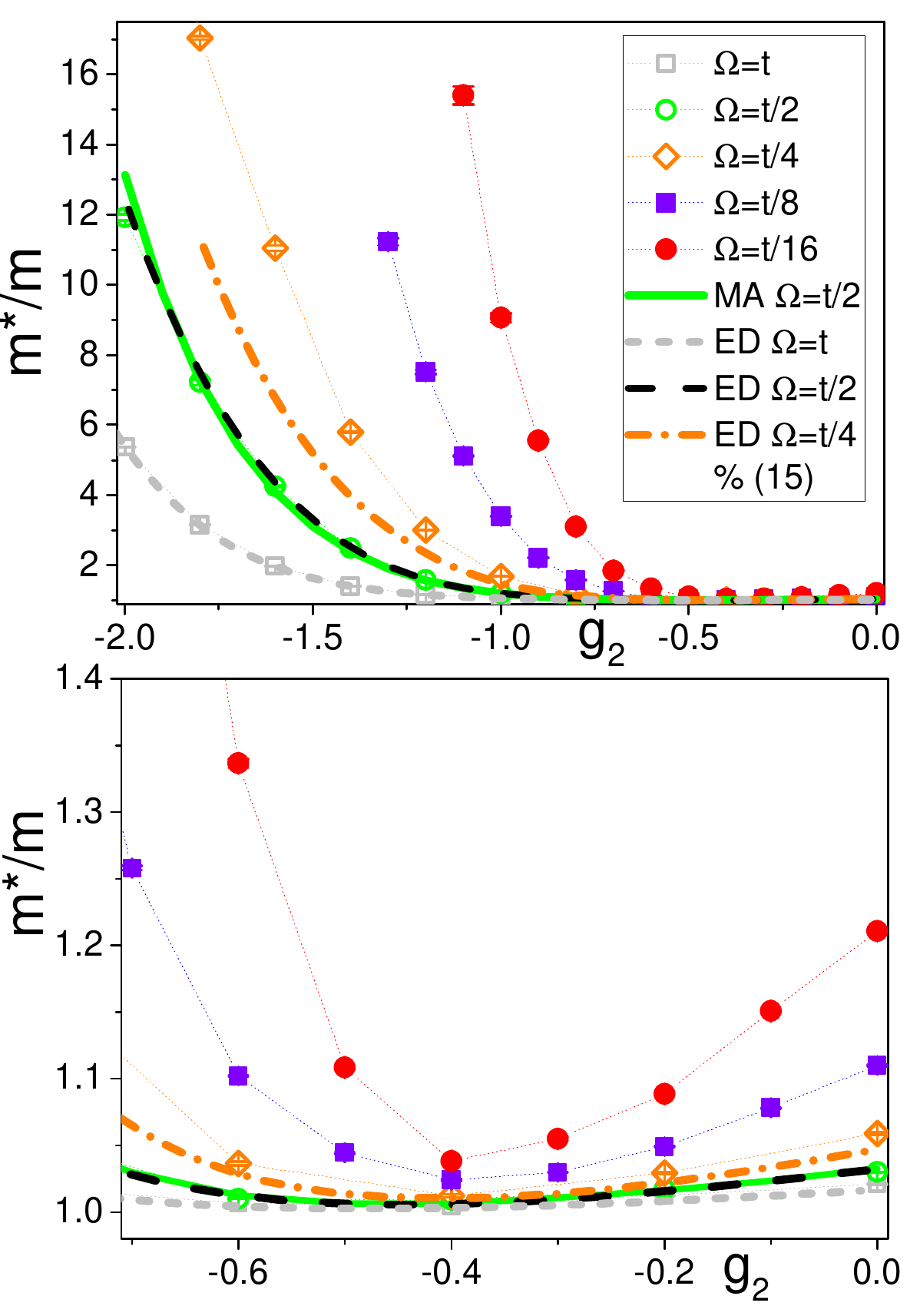}
    \caption{Effective mass renormalization $m^*/m$. For notations, see the caption of Fig.~\ref{fig:e}. Upper panel data are shown over a broad range of $g_2/t$ variation. Lower panel is focusing on the 
    small $g_2$ region.}
    \label{fig:m_be_ed}
\end{figure}
Fig.~\ref{fig:m_be_ed} presents data for the effective mass renormalization, $m^{*}/m$, where $m=1/(2t)$ is the bare mass. 
ED and MA methods provide accurate benchmarks 
for cases with relatively large phonon frequency $\Omega$
where they work best. This is no longer the case in 
the adiabatic regime due to known limitations of existing techniques, and for $\Omega \leq t/4$ we clearly observe large differences in results, especially for the $Z$ factor and effective mass.

\begin{figure}
    \centering
    \includegraphics[width=0.99\linewidth]{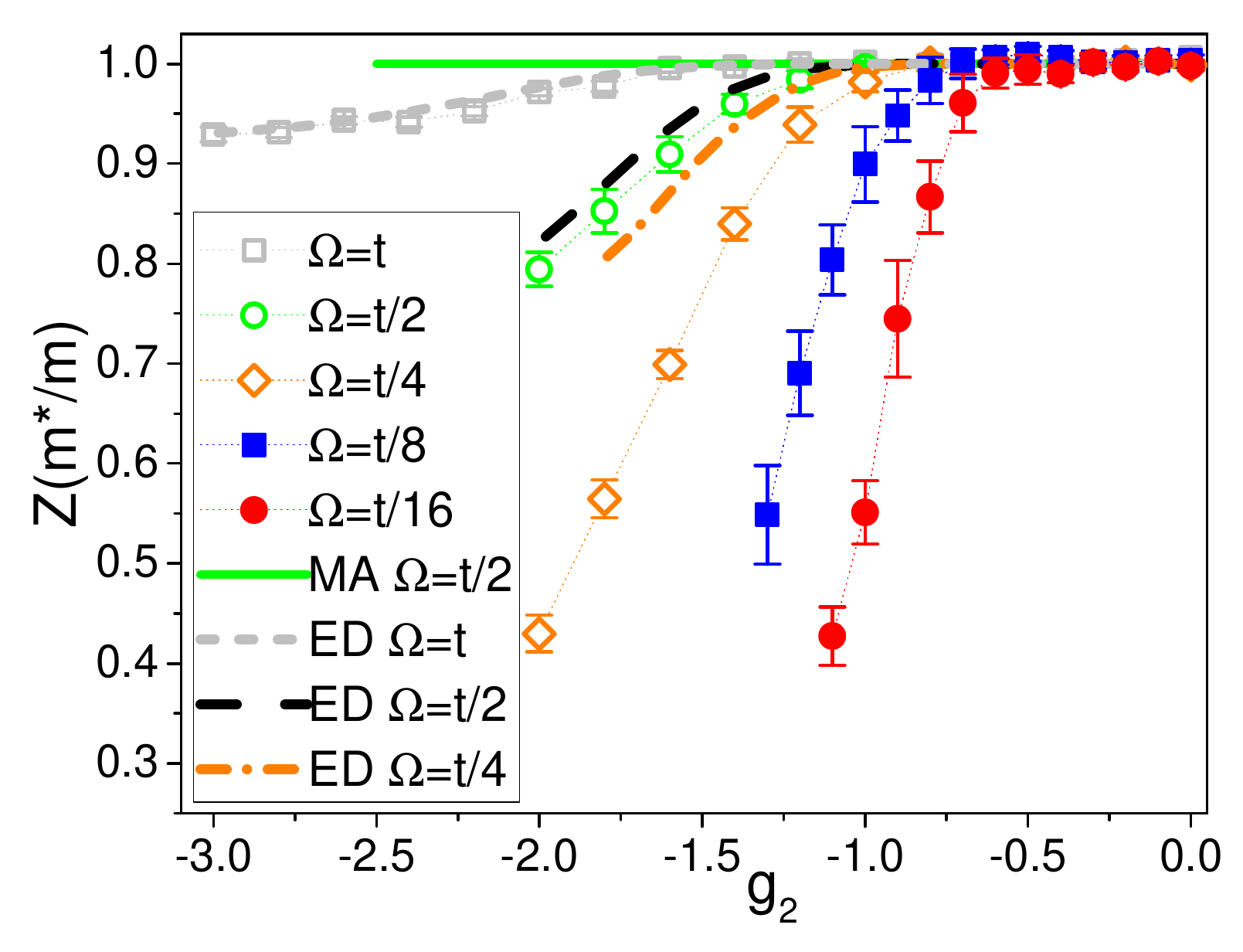}
    \caption{$Z(m^*/m)$ product as a function of $g_2/t$. For notations, see the caption of Fig.~\ref{fig:e}.}
    \label{fig:zm}
\end{figure}
 
\begin{figure}
    \centering
    \includegraphics[width=0.99\linewidth]{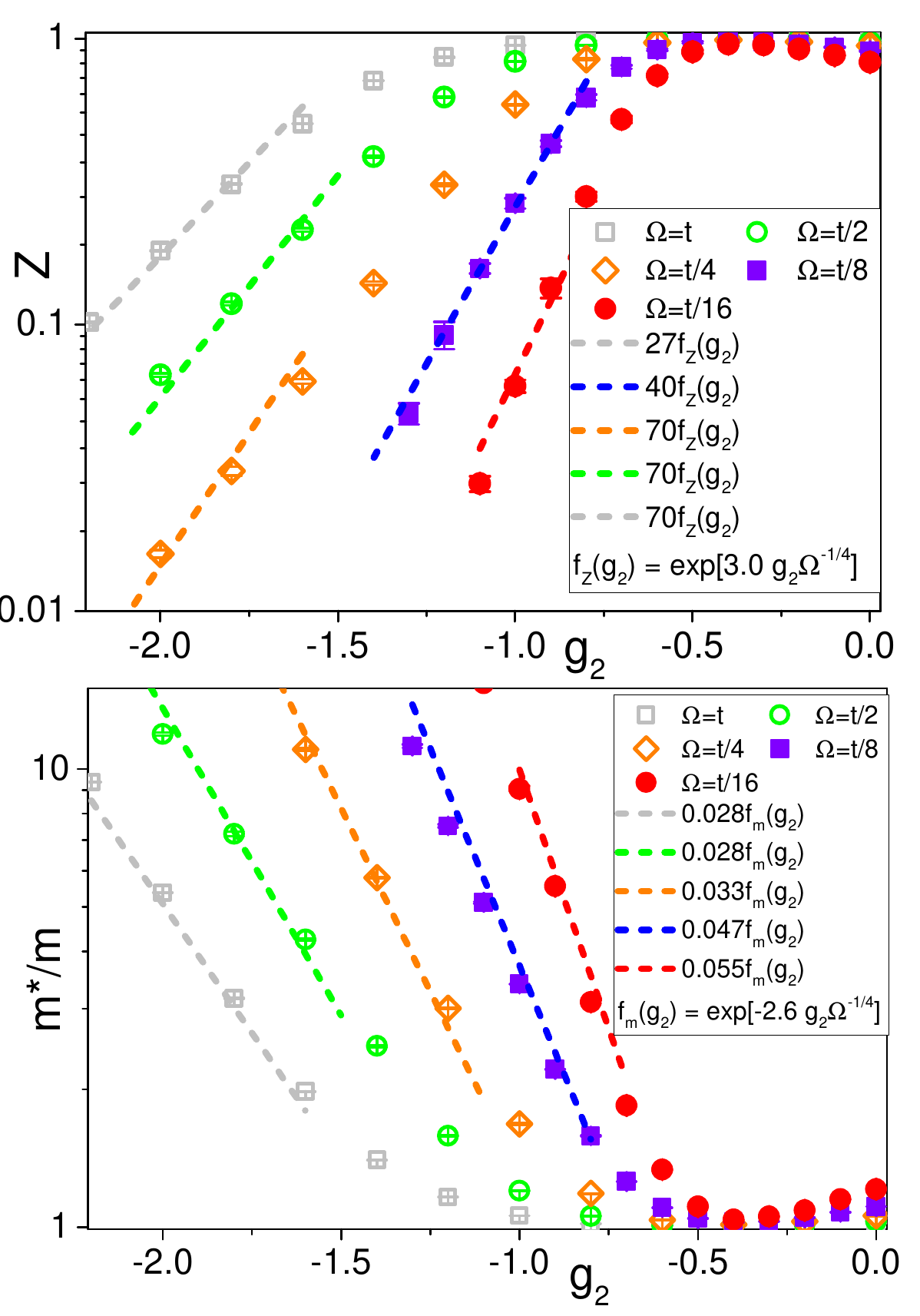}
    \caption{Universal exponential scaling of the quasiparticle weight and effective mass in the intermediate $g_2$ region, see text and
    figure legend. Notations are identical to those in Fig.~\ref{fig:e}.}
    \label{fig:scale}
\end{figure}

We start by comparing our approximation-free data with results of the pioneering MA study \cite{AdolphsEPL, AdolphsPRB, Adolphs2} at $\Omega=t/2$ shown by the solid green lines in Figs.~\ref{fig:e}, \ref{fig:z_be_ed}, and \ref{fig:m_be_ed}. We find excellent agreement for the polaron 
energy, see Fig.~\ref{fig:e}, and the first impression 
is that there is also good agreement for the quasiparticle weight (Fig.~\ref{fig:z_be_ed}) and mass renormalization (Fig.~\ref{fig:m_be_ed}), but differences may be hidden
in the strong dependence of these quantities on $g_2$.
Larger discrepancy for $Z$ and $m^*/m$ is expected because the MA technique \cite{AdolphsEPL, AdolphsPRB, Adolphs2} is based on a local treatment of the self-energy and predicts that
\begin{equation}
Z(m^{*}/m) \equiv 1 \; .
\label{locsep}
\end{equation}
It is clear from Fig.~\ref{fig:zm} that 
$Z(m^{*}/m)$ deviates from unity significantly in the strong coupling regime for all values of $\Omega/t$.
Generally, the self-energy for single polaron is always nonlocal in some parameter regimes even for local linear Holstein model \cite{ManyPo}.

Next, we compare our results with ED calculations \cite{BTB,B_P},
which do not assume locality of the self-energy but rely instead 
on a truncation of the Hilbert space. We find perfect agreement 
for all calculated properties at $\Omega=1$ (dotted gray lines in Figs.~\ref{fig:e}, \ref{fig:z_be_ed}, and \ref{fig:m_be_ed}), including the $Z(m^{*}/m)$ product, 
see Fig.~\ref{fig:zm}. 
At $\Omega/2=1/2$, the energy, $Z$-factor, and effective mass data 
are also in good agreement with ED calculations 
(dashed black lines in Figs.~\ref{fig:e}, \ref{fig:z_be_ed}, and \ref{fig:m_be_ed}), but systematic bias coming from 
the Hilbert space truncation becomes visible for $Z(m^{*}/m)$
(Fig.~\ref{fig:zm}). Finally, for $\Omega=t/4$ all properties 
computed by the ED method (dash-dot orange lines) deviate from 
our results significantly. This discrepancy originates from the necessity to keep a large but finite number of basis states in ED calculations. As the adiabatic limit (soft lattice vibrations) 
is approached, the number of relevant basis states grows 
too rapidly for the ED method to remain computationally feasible.  

It is expected that in the strong coupling limit both $Z$ and $m^*$
are exponential functions of $g_2$ as a direct consequence of the exponentially vanishing overlap between the vibrational state corresponding to empty and occupied sites. This expectation is confirmed by Fig.~\ref{fig:scale}, where exponential dependence on $g_2$ is observed for all values of $\Omega$. More precisely 
all data for $Z$ can be fit to 
$f_z = \exp \left[ \xi_Z g_2 \Omega^{-1/4} \right] $, 
and all data for $m^*/m$ can be fit to 
$f_z = \exp \left[ -\xi_m g_2 \Omega^{-1/4} \right] $,
with $\Omega$-independent $\xi_Z =3.0$ and $\xi_m =2.6$.
The scaling with frequency $\Omega$ is surprising and cannot be reproduced by simple considerations.  
Note, these exponential laws are approximately valid for more than 
an order of magnitude variation in frequency: $1/16 \le \Omega \le 1$.    
\begin{figure}
    \centering
    \includegraphics[width=0.99\linewidth]{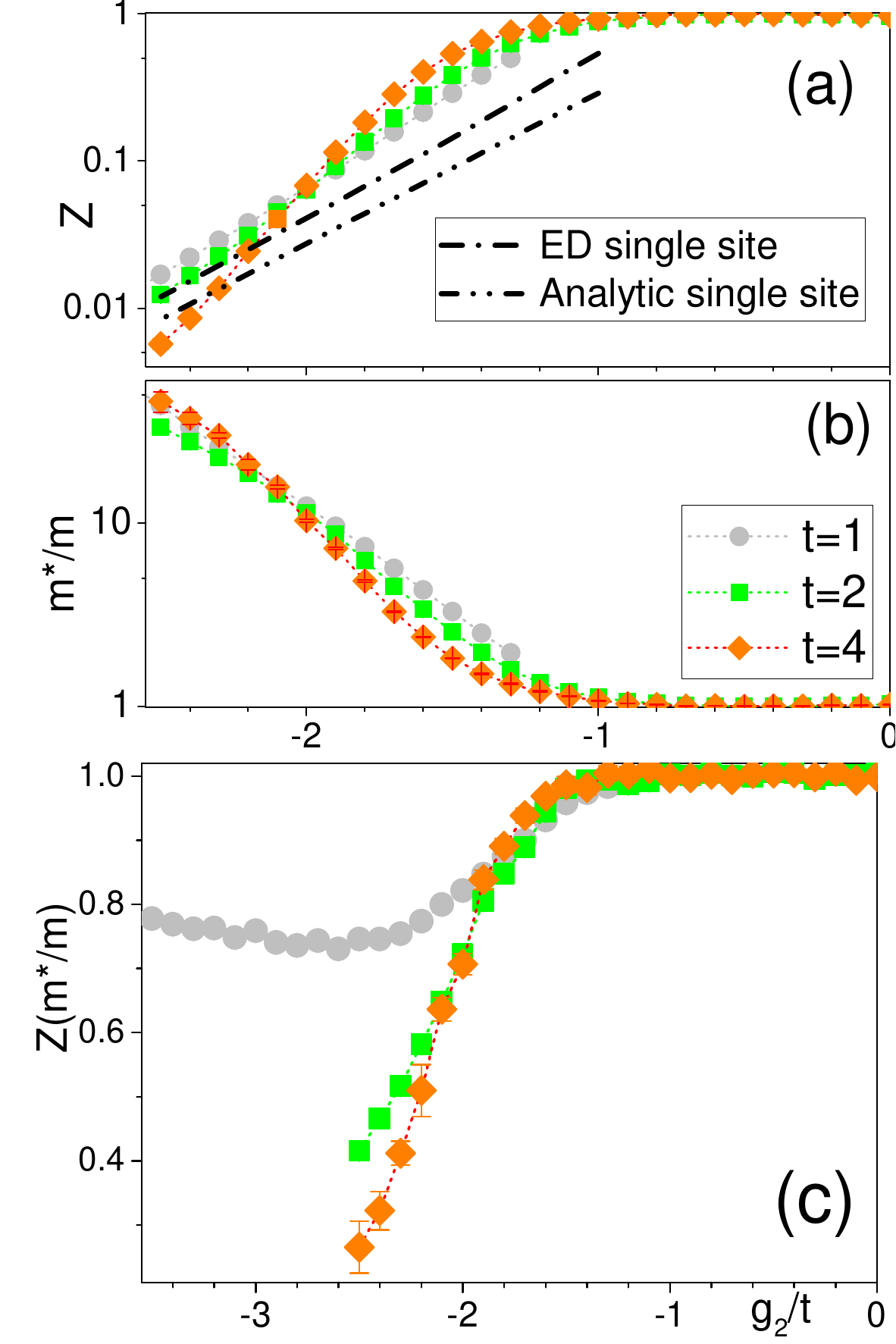}
    \caption{Polaron properties as functions of $g_2$ at $\bar\Omega=t/2$ and $\bar g_4=t/10$ for various values of $\gamma$.
    The dash-dot line is the exact asymptotic behavior of $Z$ 
    in the single site problem obtained with ED technique. The 
    dash-double dot line is the asymptotic analytic behavior (\ref{Z_analytic}).}
    \label{fig:scal_t}
\end{figure}

\subsection{Ground state properties at different hopping}
\label{fixed_omega}

In this subsection we explore how system properties evolve
when we change the hopping amplitude. In the previous section
we  used $t$ as the unit of energy and worked with the parameter set $t, \Omega, g_2, g_4$ with fixed $g_4$. 
If we change hopping by a factor of $\gamma$ (in this subsection we consider $\gamma=1,2,4$), while keeping the phonon frequency and quartic coupling fixed, we will study a model with parameter set 
\begin{equation}
\left\{ \gamma t, \bar\Omega, \bar g_2, \bar g_4 \right\}_{\bar g_2, \gamma} \; .
\label{parset}
\end{equation}
Since the choice of the energy unit is arbitrary, 
we can equivalently consider the set 
$$
\left\{ t, \Omega=\frac{\bar\Omega}{\gamma}, g_2=
\frac{\bar g_2}{\gamma}, g_4=\frac{\bar g_4}{\gamma} \right\}_{g_2, \gamma}.
$$
In what follows, we fix $\bar\Omega=t/2$ and $\bar g_4=t/10$. 
The reason for considering this parameter scaling is to 
compare all curves to the unique single-site limit, 
see Fig.~\ref{fig:scal_t}, since, at each $g_2$, all parameters of the local Hamiltonian (\ref{loc1}) are kept fixed. 

It is well known that for linear EPI (no matter whether the coupling is to optical \cite{Alvermann} or acoustic \cite{HahnAc} modes) the crossover to the strong coupling regime at $\lambda=g_1^2/(2t\Omega)\sim 1$ 
is sharper for smaller $\Omega/t$. We observe a similar trend, see
Fig.~\ref{fig:scal_t}, but the effect is much less pronounced.   

The other difference from the linear case is how quickly the strong-coupling case is adequately described by the asymptotic AL, $g_2 \ll -1$, when the anharmonic potential can be approximated by two degenerate harmonic wells at local vibrational coordinates $\pm x_0$ (\ref{x00}) and atomic wavefunctions, by 
a superposition of two shifted Gaussians.
In this limit (see Appendix \ref{assyan})
\begin{equation}
    Z_0(t=0) = 2 \times \frac{2\sqrt{R}}{1+R}
    \exp\left( -\frac{R}{1+R} \, z_0^2 \right) \; , 
\label{Z_analytic}    
\end{equation}
where
\begin{equation}    
    z_0^2 = x_0^2 \Omega = \frac{-4g_2-\omega}{16g_4} \; ,
\label{z0z0z0}    
\end{equation}
and $R$ is given by Eq.~(\ref{RRR}).
As one can see in Fig.~\ref{fig:scal_t} our data are still far from the single-site asymptotic behavior despite $Z$ and $m^*$ changing 
by nearly two orders of magnitude. [Comparison with SSED result in this limit shows that the asymptotic behavior is reached only at 
$|g_2| > 5$, see Fig.~\ref{fig:el1}.]

Finally, in the adiabatic case, starting from $\Omega/t=1/4$, we observe
that the $Z$ factor drops below the AL prediction. This effect
was never reported in the linear case. For positive quadratic coupling,
$g_2 \gg 1$ and $g_4=0$, it signals the formation of the extended
soliton (or ``self-trapped") state \cite{soliton}. 
It appears that for the double-well potential polarons can also
form extended self-trapped states (finite volume with large lattice distortion moving as a whole) before reaching AL limit.  

The above considerations show that the physics is reduced to a single-site problem for the asymptotically large coupling limit. Therefore, the self-energy in this regime must become local, and the relation (\ref{locsep}) has to be valid again. This is confirmed by behavior of $Z(m^*/m)$ at large values of $|g_2|$, see Fig.~\ref{fig:scal_t}c.

\begin{figure}
    \centering
    \includegraphics[width=0.99\linewidth]{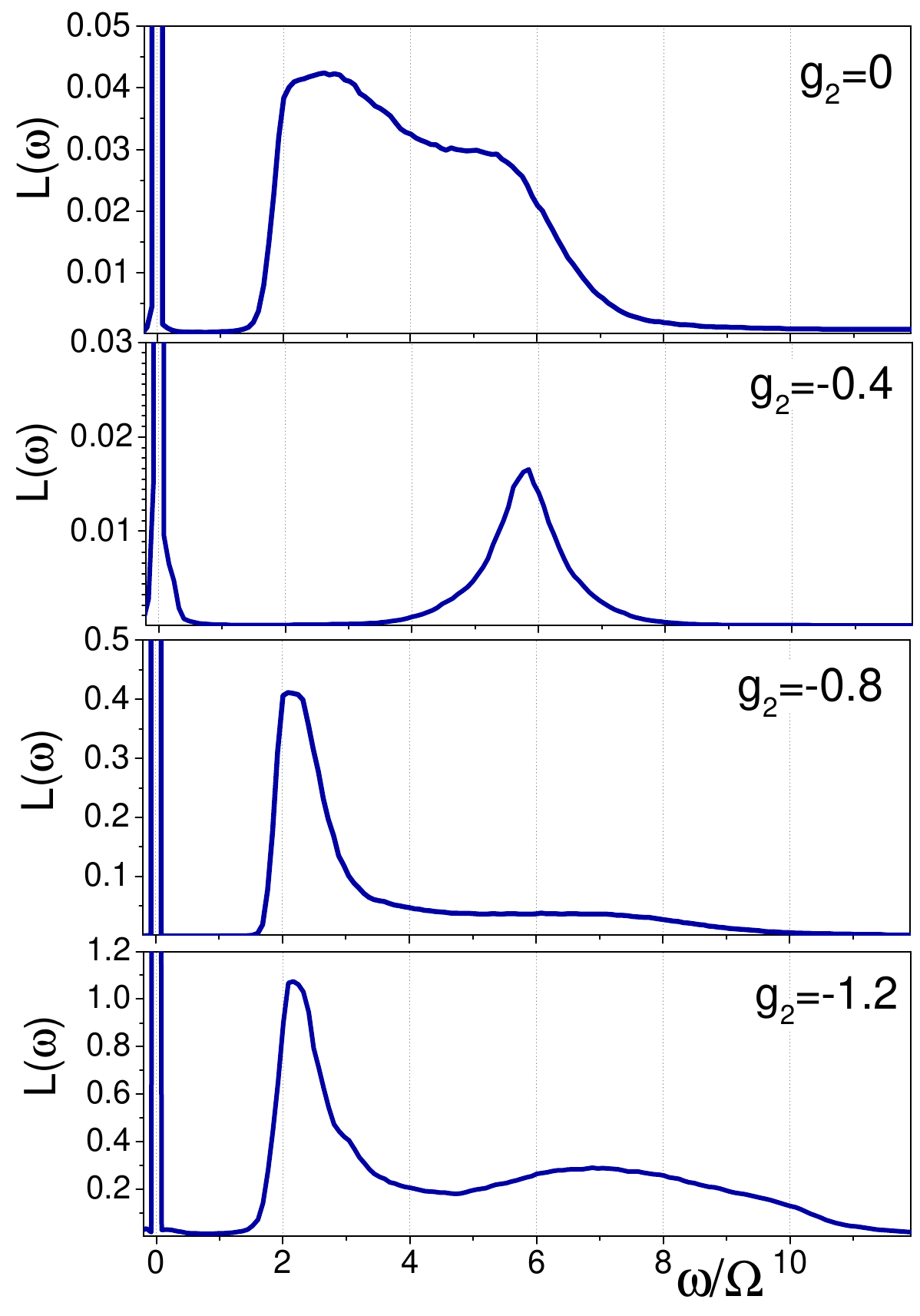}
    \caption{Lehmann spectral function for $t=1$, $g_4=0.1$, $\Omega=t/4$, and different values of $g_2$. Frequency is 
    counted from the ground state energy in units of $\Omega$.}
    \label{fig:lehman}
\end{figure}

\subsection{Spectral function and the structure of the phonon cloud} 
\label{spectral}

In this subsection, we investigate the polaron Lehmann 
spectral function $A(\omega)$, obtained by analytic 
continuation of the equation
\begin{equation}
G(\tau) = \int_0^{\infty} d\omega \; e^{-\omega\tau} A(\omega) \,,
\label{ac}    
\end{equation}
using the Stochastic Optimization with Consistent Constraints \cite{MPSS, Julich, SOCC} method. 

As intuitively expected, the spectra presented in Fig.~\ref{fig:lehman}
for $g_2=0$, $g_2=-0.8$ and $g_2=-1.2$ show a clear 
threshold for excited states at $\omega= 2\Omega$. Indeed, in perturbation theory both $x^2$- and $x^4$-terms can change vibrational states only by zero, two, four, or any other even number of quanta.
Therefore, one expects, in close analogy with 
well-understood linear EPI cases \cite{MPSS, OC2003, FC2006, OCHol_Ber}, that the corresponding low-energy threshold for incoherent quadratic and quartic processes is $2\Omega$. Moreover, since the spectral weight at $\omega=4\Omega$ for $g_2=-1.2$ is larger than that for $g_2=-0.8$, we conclude that for our parameters 
the main contribution at this frequency originates 
from the $x^2$ terms at higher expansion order. 
Obviously, for $g_2=0$ the spectral continuum is entirely controlled by the $x^4$-terms.   

Given these considerations, the structure of the Lehmann spectral function for $g_2=-0.4$ appears to violate the rule because the 
low-energy threshold is observed at $4\Omega$. 
However, as already noticed in Ref.~\cite{AdolphsPRB}, for negative $g_2$ there is destructive ``interference" between quadratic and quartic interactions leading to the effective suppression of the total EPI strength and $Z \approx 1$. 
In the lowest-order perturbation theory, this destructive 
interference argument predicts that the spectral weight goes to zero at $\omega=2\Omega$ for $g_2=-6g_4$, or $g_2=-0.6t$ for $g_4=0.1t$
considered here. Due to the strong competition between $g_2$ and $g_4$, we name this specific scenario as \emph{quantum interplay} regime.

\begin{figure}
    \centering
    \includegraphics[width=0.99\linewidth]{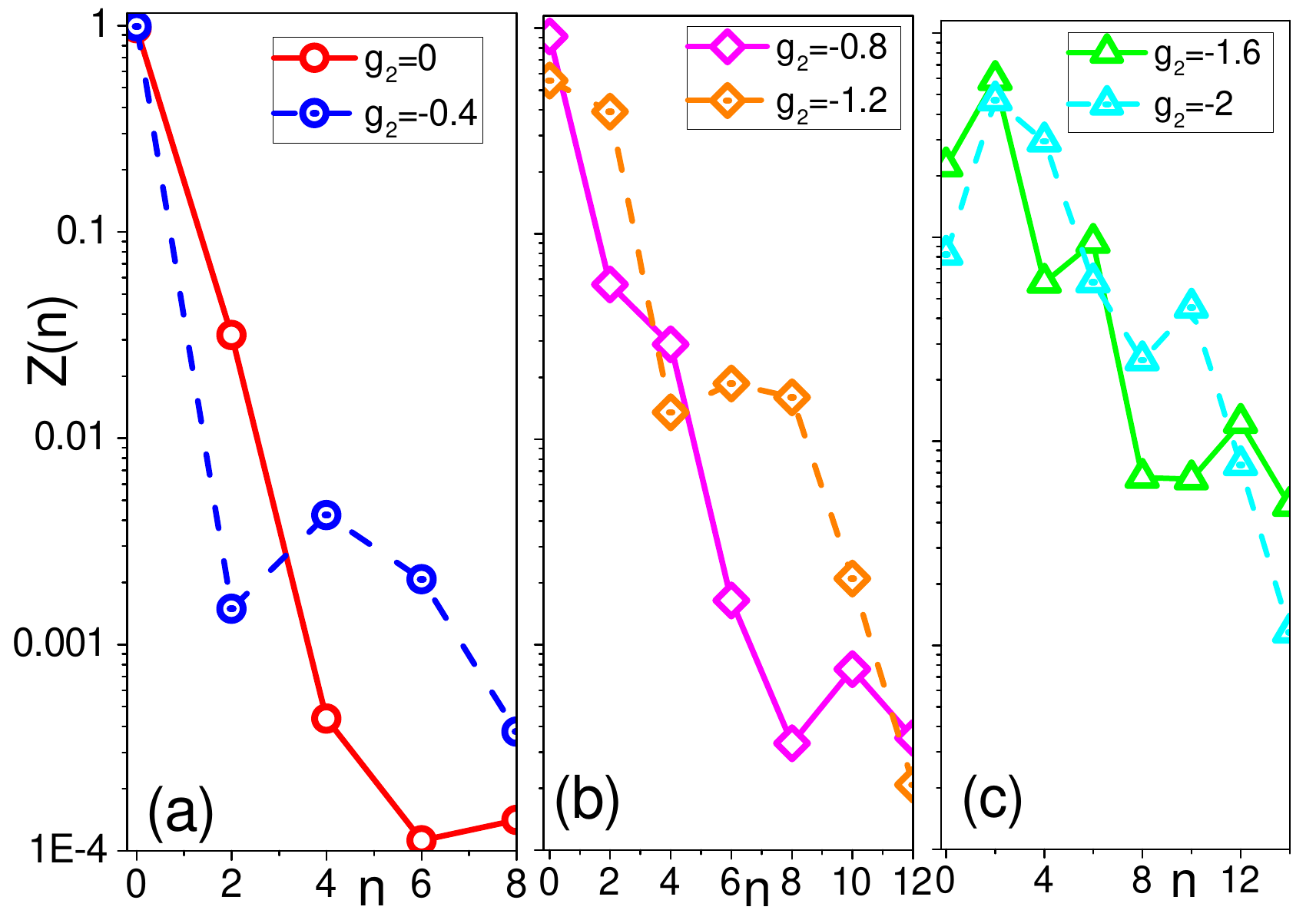}
    \caption{Partial weights of states with a given number of phonons in the polaron cloud ($\sum_n Z(n)=1$) for $g_4=0.1t$, $\Omega=t/4$, and different values of $g_2$.}
    \label{fig:wave_fun}
\end{figure}

A more precise treatment of the observed behavior is provided by ED studies of the partial weights $Z(n)$ of states with $n$ phonons in the exact solution. 
We find that the weight of 2-phonon states is significantly suppressed at $g_2=-0.4$, see Fig.~\ref{fig:wave_fun}(a), to such a degree that the numerical analytic continuation can not resolve it. 
The other interesting feature of the phonon cloud structure, which has never been observed before in systems with nonlinear EPI,
is the oscillatory dependence of $Z(n)$ on $n$ and $g_2$ at large coupling $|g_2| > 1$, see Fig. \ref{fig:wave_fun}(b,c).

\section{Conclusions}
\label{conclu}

In this work, we developed the Numeric X-propagators technique for accurately solving a broad class of polaron models without approximations. This approach can treat
(i) arbitrary coupling of the inter-site hopping amplitude magnitude to the local coordinates of atoms on sites $i$, $j$, and 
the bond $\langle ij \rangle$ between them, i.e.  
$t_{ij}(x_i,x_j, x_{ij}) = t \exp \{ S(x_i,x_j, x_{ij}) \} $
with an arbitrary non-pathological function $S$ \cite{x-representation22};
(ii) arbitrary coupling between the carrier density and the on-site atomic vibration: $n_i V_{\rm ep}(x_i)$.
The method itself is a hybrid of the precise numeric tabulation of the atomic coordinate X-propagator in imaginary time (either by Feynman path-integral or exact diagonalization techniques) with the lattice path integral
for a single electron. The latter is ``dressed" by the X-propagators and simulated by the Diagrammatic Monte Carlo technique. 
An approximation-free solution for the electron Green's function in imaginary time allows one to obtain unbiased values for the polaron ground state energy, quasiparticle weight, and effective mass. 
Numeric analytic continuation of the Green's function provides access to the
Lehmann spectral function and properties of excited polaron states, which are
measurable through angle-resolved photoemission experiments (ARPES).

As an application, we studied a particular case of the double-well potential coupling, which is considered prototypical for several classes of materials. We compared our results to earlier studies \cite{AdolphsPRB, Adolphs2} of the same model within the Momentum Average approximation framework. We find that for cases with relatively large phonon frequency, $\Omega\geq t/2$, this approximation provides a reasonable description of the polaron energy and quasiparticle weight, but the key property of the Momentum Average approximation, $(m^*/m)Z \equiv 1$, fails. The exact diagonalization technique provides a more accurate description in the same parameter range. However, both Momentum Average and Exact Diagonalization methods start to fail in the adiabatic regime $\Omega \le t/4$, leaving our technique as
the only unbiased approach suitable for dealing with realistic materials. 
We extended our calculations to the deep adiabatic regime 
$\Omega=t/16$, where the phonon frequency is almost two orders of magnitude smaller than the electron bandwidth.   

Given that the simplest possible double-well model has four parameters, 
and its properties depend on three dimensionless ratios, we fixed 
the quartic coupling $g_4 >0$, and investigated system properties for
different adiabatic ratios $\Omega/t$ as functions of quadratic coupling $g_2$. 
We identified three qualitatively different parameter regimes:
(i) The quantum interplay regime, $-g_2 \sim 4 g_4$, where
 quadratic and quartic terms compete and the separation between the
 two minima of the potential $V(x)$ on the occupied site
 is comparable to the ionic zero-point vibration. In this regime, the
destructive interference between quadratic and quartic terms may result in
near complete suppression of the effective electron-phonon interaction, 
leading to $Z \approx 1$, $m^*/m \approx 1$, and even to the disappearance of the 
spectral continuum threshold at excited energy $2\Omega$.
(ii) The moderately strong coupling regime, $-g_2 < 4 g_4$, in which 
the quasiparticle weight and effective mass are described by exponential laws
$Z \propto \exp[\xi_Z g_2 \Omega^{-1/4}]$ and $m^*/m \propto \exp[- \xi_m g_2 \Omega^{-1/4}]$. 
(iii) An extreme strong coupling regime at very large, $|g_2| > 100 \, |g_4|$, values of the quadratic coupling constant. This regime corresponds to the asymptotic 
single-site double-well state, with exponentially small $Z$ and $m/m^*$, and which is reached only at very large, $|g_2| > 100 \, |g_4|$, values of the quadratic coupling constant.

In this work, we have shown that the newly introduced Numeric X-propagators technique is able to accurately solve a polaron model which features strongly anharmonic potentials that exhibit a double-well structure. This is one of the main ingredients required to investigate charge transport at low-doping concentrations in quantum paraelectric and ferroelectric materials such as $\text{SrTiO}_3$ and $\text{BaTiO}_3$. This work lays the foundation for future studies with material-specific potential profiles obtained from ab-initio calculations.

Moreover, it is important to emphasize that the numerical X-propagator technique is not restricted to the case of dispersionless phonons considered in this study.
Indeed, a straightforward generalization of this method to the case of dispersive phonons is possible, making our approach a fundamental tool for approximation-free studies of transport and optical properties \cite{HolstMobil, FrohMobil} in a wide range of realistic materials. In particular, the rigorous treatment of strong phonon dispersion becomes especially essential for quantum paraelectric and ferroelectric materials, halide perovskites, and anharmonic superconductors~\cite{Setty2024}, where the phonon dispersion considerably exceeds the minimal phonon frequency \cite{PhoDisp}.    

\section{Acknowledgments} 

This work was supported by the Croatian Science Foundation under projects number IP-2024-05-2406, IP-2022-10-9423, and IP-2022-10-3382. 
O.S.B. acknowledges support of the project FrustKor financed by the EU through the National Recovery and Resilience Plan 2021-2026 (NRPP).
S.R. and C.F. acknowledge support from the joint Austrian Science Fund (FWF)—FWO projects I4506 and PIN5456724. 
N.P. acknowledges support from the National Science Foundation under Grant No. DMR-2335904.
N.N. was supported by JSPS KAKENHI Grant Numbers 24H00197, 24H02231, 24K00583, and
by the RIKEN TRIP initiative.
Most computations were carried out on the Vienna Scientific Cluster (VSC) and the computational facilities of RIKEN Center for Emergent Matter Science (CEMS).

\appendix

\input{trotter-suzuki}

\input{ed}

\input{direct-estimators}

\input{Z_asymptotic}

\bibliography{bibliography}

\end{document}

%% file: trotter-suzuki.tex
\section{Calculation of the oscillator propagator with arbitrary electron-phonon potential by Trotter-Suzuki decomposition}
\label{xprop-app}

This section describes the method of calculation of the X-propagator in the case of a general electron-phonon potential $V_{ep}(X)$ which, in contrast to the SSED method, 
does not require the representation of the interaction potential in the form (\ref{loc1}) or (\ref{loc4}). 

The propagator (\ref{xprop_dep}) can be presented as a path integral
\begin{eqnarray}
&{\tilde U}(x(\tau_f),x(\tau_i),\tau_f-\tau_i) = A(\tau) \;  
e^{\Omega\tau/2} \times
\nonumber \;\;\;\;\;\;\;\;\;\; & \\
& \int\limits_{\rm all~ paths} \exp \left\{ - \int\limits_{\tau_i}^{\tau_f} 
\left[ \frac{1}{2} \left( \partial x / \partial \tau \right)^2
+ V(x) \right]
\right\}  \; , & 
\label{path}    
\end{eqnarray}
where $A(\tau)$ is a normalization constant. The exponent $e^{\Omega\tau/2}$
is added to provide $\tau$-independent asymptotic behavior for $\tau \to \infty$.

\subsection{Analytic expressions for specific potentials}
\label{analexp}

The explicit form of the X-propagator is known when $V_{ep}(x)=0$. It reads \cite{Moriconi}   
\begin{equation}
U(x,y,\tau)  = \sqrt{\frac{\Omega}{2\pi \sinh[\Omega\tau]}} e^{\frac{1}{2}\Omega\tau - Q(x,y,\tau)}  \; , 
\label{FPP1}
\end{equation} 
where
\begin{equation}
Q(x,y,\tau) = \Omega\frac{\cosh[\Omega\tau](x^2+y^2)-2xy}{2 \sinh[\Omega\tau]} \;  .
\label{FPP2}
\end{equation} 
An important property of the X-propagator is the limiting behavior
\begin{equation}
U(x,y,\tau \to \infty) = U_0(x)U_0(y) \; ,    
\end{equation}
where 
\begin{equation}
U_0(x) = (\Omega/\pi)^{1/4} e^{-\Omega x^2 / 2} \; .
\label{u0}    
\end{equation}

Also, when the expansion of $V_{\rm ep}(x)$ is limited to linear ($g_1 \ne 0$) and quadratic ($g_2 \ne 0$) terms, one can transfer the potential (\ref{loc3}) to
\begin{equation}
V(x) =  \frac{\Omega_{\rm at}^2}{2}    
\left( x + \xi_{\rm sh} \right)^2 + \Delta_{\rm sh} + 
\frac{\Omega_{\rm at}}{2}, 
\label{apts1}
\end{equation}
where 
\begin{equation}
\Omega_{\rm at} = \Omega \sqrt{1 + n \frac{4g_2}{\Omega}} \; ,
\label{HinXtrFin}
\end{equation}
\begin{equation}
\xi_{\rm sh} = \frac{g_1}{\Omega_{\rm at}^2} \sqrt{2\Omega} \; , 
\label{shishi}
\end{equation}
and 
\begin{equation}
\Delta_{\rm sh} = - \frac{g_1^2}{\Omega_{\rm at}^2} \Omega \; .
\label{HinXtrFin2}
\end{equation}
In this case, the X-propagator is  
\begin{equation}
\tilde{\bf U}(x,y,\tau)  = \sqrt{\frac{\Omega_{\rm at}}{2\pi \sinh[\Omega_{\rm at}\tau]}} e^{\frac{1}{2}\Omega\tau - \tilde{Q}(x,y,\tau) +  \Delta_{\rm sh}}  \; , 
\label{FPP1a}
\end{equation}  
where
\begin{equation}
\tilde{Q}(x,y,\tau) = \Omega_{\rm at} \frac{\cosh[\Omega_{\rm at}\tau](\bar{x}^2+\bar{y}^2)-2\bar{x}\bar{y}}{2 \sinh[\Omega_{\rm at}\tau]} \;  
\label{FPP2a}
\end{equation} 
and
\begin{equation}
\bar{x}=x+ \xi_{\rm sh}, \; \bar{y}=y+ \xi_{\rm sh}  \; .
\label{FPP3}
\end{equation}

Thus, if an arbitrary potential $V(x)$ on the $(x,x')$ 
interval is Taylor expanded around point $\bar{x}=(x+x')/2$:
\begin{equation}
V(x) \approx 
V(\bar{x}) + V'(\bar{x}) (x-\bar{x}) + V''(\bar{x})(x-\bar{x})^2/2  \; ,
\label{TEV1}
\end{equation}
one can use Eqs.~(\ref{FPP1a})-(\ref{FPP3}) to obtain the most
accurate analytic solution for the propagator
$P(x,x',\delta \tau)$. 

\subsection{Path-integral representation for an arbitrary potential: Monte Carlo updates}
\label{TroSuz}

For a concise description of updates, let us represent the propagator between any left $i$ and right $i+1$ points (\ref{path3}) as 
\begin{equation}
P(x_i,x_{i+1},\delta\tau) = \exp\{-\Theta[x_i,x_{i+1}]\} .
\label{repru}    
\end{equation}
Then, four updates will ensure ergodic sampling of the configuration space: changing $x_i$ at intermediate points, changing initial (final) $x_i$ ($x_{f}$) coordinates, adding and removing one slice at the end.

\subsection{Updates}

\textit{Intermediate point update.} In this case, we chose any intermediate time slice $i$ and propose to change $x_i$ to $x_i \pm z$ 
by selecting $z$ from the exponential distribution   
\begin{equation}
W(z) = \kappa e^{-\kappa z} \;, 
\label{exse2}    
\end{equation}
where $\kappa$ is uniformly seeded in the range $1/\Omega m < \kappa < m/\Omega$ with $m \approx 3$.   
The Metropolis algorithm ratio is 
\begin{align}
r = \exp\{&-\theta[x_{i-1},x_{i}\pm z ]-\theta[x_{i}\pm z,x_{i+1}] 
\nonumber \\
  & + \theta[x_{i-1},x_i]+\theta[x_i,x_{i+1}]\} \; .
\label{exse3}    
\end{align} 

\textit{Initial (final) point update.}
Now we select either the initial or final point
and propose to update its oscillator coordinates by exactly the same protocol as for intermediate points. The Metropolis ratio is now (for final point)
\begin{equation}
r = \exp\{-\theta[x_{f-1},x_{f} \pm z ] + 
\theta[x_{f-1},x_{f}]\} \; .
\label{exse4}    
\end{equation} 

\textit{Adding (removing) slice update.} We propose to add one slice at the end with $x_{f+1} = x_f$. Again, we generate $z$ from Eq.~(\ref{exse2}) and propose to change $x_f$ to $x_f \pm z$. The Metropolis ratio is now
\begin{equation}
r = \frac{\exp\{-\theta[x_{f-1},x_f \pm z]-\theta[x_f \pm z,x_f] + \theta[x_{f-1},x_f]\}}
{(\kappa/2) \exp(-\kappa z)}\; .
\label{exse5}    
\end{equation} 
In the removal case, the ratio is the reciprocal.

%% file: ed.tex
\section{Calculation of the oscillator propagator 
by Single-Site Exact Diagonalization (SSED)}
\label{ededed}

Consider Eq.~(\ref{loc1}) at a single site
\begin{equation}
    \hat H = \Omega \, \hat b^\dag \hat b + \sum_{i=1,2,4} g_i (\hat b + \hat b^\dag)^i \, .
\end{equation}
Its matrix elements in the phonon number basis are straightforwardly calculated. For $m \ge n$,
\begin{align}
    \bra{m} \hat H_1 \ket{n} = g_1 \cdot \delta(m, n-1) \cdot \sqrt{n} \, ,
\end{align}
\begin{align}
     \bra{m} \hat H_2 \ket{n} = g_2 \cdot
     (&\delta(m, n-2) \cdot \sqrt{n(n-1)} \nonumber \\
     +&\delta(m, n) \cdot (1 + 2n)) \, ,
\end{align}
\begin{align}
     \bra{m} \hat H_4 \ket{n} = g_4 \cdot
     (&\delta(m, n-4) \cdot \sqrt{n(n-1)(n-2)(n-3)} \nonumber \\
     +&\delta(m, n-2) \cdot 2(2n-1)\sqrt{n(n-1)} \nonumber \\
     +&\delta(m, n) \cdot 3 (2n^2 + 2n + 1)) \, .
\end{align}
This matrix is real and symmetric, so it can be diagonalized with real eigenvalues and eigenvectors. The energy eigenstates can be labeled by a discrete quantum number $\xi$. The complete basis is composed of an infinite number of them, $\{\ket{\xi}\}_{\xi=0}^\infty$. In order to perform the diagonalization with a computer, we must impose an upper cutoff in the number of phonon states that we consider. When this cutoff is made large enough, we recover the exact eigenstates.

The propagator may then be expressed in terms of the eigenvalues and eigenvectors resulting from the diagonalization
\begin{gather}\label{u_ed_def}
    U(y,x,\tau) \equiv \bra{y} e^{-\tau \hat H} \ket{x} = \\
    \sum_{\xi} \braket{y|\xi} \braket{\xi|x} e^{-\tau E_\xi} = \\ \label{u_ed}
    \sum_{\xi} \sum_{m,n} \braket{y|n} \braket{n|\xi} \braket{\xi|m} \braket{m|x} e^{-\tau E_\xi} \, ,
\end{gather}
where $\braket{y|n}$ is the Hermite function $\psi_n(y)$ and $\braket{m|x}$ is the complex conjugate of the Hermite function $\psi_m(x)$. Because Hermite functions are real, $\psi^*_m(x) = \psi_m(x)$.
$\braket{n|\xi}$ and $\braket{\xi|m}$ are the expansion coefficients of the eigenvectors of $\hat H$ in the phonon number basis. Since the eigenvectors are real, $\braket{\xi|m} = \braket{m|\xi}$.
$E_\xi$ are the eigenvalues obtained by the diagonalization of $\hat H$.

The evaluation of the quantity in Eq.~(\ref{u_ed}) is too expensive to be performed at each Monte Carlo step. For this reason, the propagator is precomputed in advance at a dense grid of points, and continuous values in between are obtained by linear interpolation.

The single-site Green's Function (\ref{eq:gf0_u}) is expressed in terms of Eqs.~(\ref{u_ed_def}) and (\ref{u0}):
\begin{equation}
    G^0(\tau) = \int_{-\infty}^\infty \text{d}y \int_{-\infty}^\infty \text{d}x \,
    U_0(y) U(y,x,\tau) U_0(x) \, .
\end{equation}

By the orthonormality of the Hermite polynomials $H_n(x)$ to the constant function $1$,
\begin{equation}
    \int_{-\infty}^\infty \text{d}x \, e^{-\Omega x^2} H_n(\sqrt{\Omega}x) =
    \sqrt{\frac{\pi}{\Omega}} \, \delta_{n=0} \, ,
\end{equation}
one obtains
\begin{equation}\label{ed_gf}
    G^0(\tau) = \sum_{\xi} \braket{0|\xi}^2 e^{-\tau E_\xi} \, ,
\end{equation}
so that $Z^\text{AL}_0 = \braket{0|\xi}^2$.

The SSED method has the advantage that we obtain with great precision many quantities useful for the Monte Carlo process. One is the $\tau$-derivative of Eq.~(\ref{u_ed}), which enters the energy estimator that will be described in the next appendix. Another is the integral of $G_0(\tau)$, which allows us to get the correct normalization for the measurement of the Green's function. The collected histogram is multiplied by the normalization constant
\begin{equation}
    C = \frac{N}{N_0} \, I_0 \, , \qquad
    I_0 = \int_{0}^{\tau_\text{max}} \text{d}\tau \, G^0(\tau) \, .
\end{equation}
where $N_0 / N$ is the statistics of the zeroth order diagram, and $I_0$ is obtained by integrating Eq.~(\ref{ed_gf}) term by term. This normalization method ensures better statistics than the simpler method employing the property $G(\tau = 0) = 1$ and removes its systematic error due to the size of the $\tau$-bin.

Another advantage of this method over the path integral one is the possibility of obtaining the propagator for a nonuniform grid in imaginary time, better suited for the region of small imaginary times where the Green's function changes rapidly.

%% file: direct-estimators.tex
\section{Direct estimators for the energy and the effective mass}

Alongside with the usual method which extracts ground state properties from the Green's function, we derive direct estimators for the energy and effective mass.

Both derivations follow the same strategy: equating the Green's function in the asymptotic $\tau$ limit
\begin{equation}
    G(k,\tau \to \infty) \to Z_0(k) e^{-E_0(k)\tau} = e^{-E_0(k)\tau + \ln(Z_0(k))}
\label{gf_asympt}
\end{equation}
to its general expression, transformed to position space and expanded into the perturbation series
\begin{align}\label{pert-exp}
    G(\Delta r, \tau) = \sum_{\nu} W_\nu(\Delta r, \tau) \, ,
\end{align}
where $\nu \equiv (n, \{\tau_i\}, \{x_j\})$ is a collective coordinate which includes all summation indices and integration variables of the diagram: order $n$ (number of hoppings), imaginary times $\{\tau_i\}$ and displacements $\{x_j\}$, and $W_\nu$ is the diagram weight.

\subsection{Energy estimator}

On one hand, taking the logarithmic derivative with respect to $\tau$ of (\ref{gf_asympt})
\begin{equation}
    \frac{1}{G(k,\tau)} \frac{\text{d}G(k,\tau)}{\text{d}t} = -E_0(k) \, .
\end{equation}
On the other hand, using the perturbation expansion (\ref{pert-exp})
\begin{multline}
    \frac{1}{G(k,\tau)} \frac{\text{d}G(k,\tau)}{\text{d}t} =
    \frac{\sum_{\Delta r} e^{-ik \Delta r} \frac{\text{d}G(\Delta r,\tau)}{\text{d}t}}{\sum_{\Delta r} e^{-ik \Delta r} G(\Delta r, \tau)} = \\
    \frac{\sum_{\Delta r,\nu} e^{-ik \Delta r} W_\nu(\Delta r,\tau) \frac{1}{W_\nu(\Delta r,\tau)} \frac{\text{d}W_\nu(\Delta r,\tau)}{\text{d}t}}{\sum_{\Delta r,\nu} e^{-ik \Delta r} W_\nu(\Delta r, \tau)} = \\
    \frac{\left\langle e^{-ik \Delta r} \frac{1}{W_\nu(\Delta r,\tau)} \frac{\text{d}W_\nu(\Delta r,\tau)}{\text{d}t} \right\rangle_\text{MC}}{\langle e^{-ik \Delta r} \rangle_\text{MC}} \, .
\end{multline}
In order to take the $\tau$-derivative of the diagram weight $W$ one must make explicit the dependence on $\tau$ of the internal times $\tau_i$:
\begin{equation}
    \tau_i' \equiv \tau_i / \tau \, , \qquad
    \tau_i = \tau \, \tau_i' \, .
\end{equation}
Due to the change of integration variables $\tau_i \to \tau_i'$, $\text{d}\tau_i = \tau \text{d}\tau_i'$ so the new diagram weight becomes
\begin{equation}
    W_{\nu'}(\Delta r,\tau) = \tau^n W_\nu(\Delta r,\tau) \, , \qquad
    \nu' \equiv (n, \{\tau_i'\}, \{x_j\}) \, ,
\end{equation}
\begin{equation}
    \frac{\text{d} \ln W_{\nu'}(\Delta r,\tau)}{\text{d}\tau} =
    \frac{1}{\tau} \left(n + \sum_i \frac{\text{d}\ln U(y_i,x_i,\tau_i)}{\text{d}\tau_i} \, \tau_i \right) \, .
\end{equation}
The final result for the energy estimator is
\begin{equation}
    E_0(k) =
    \frac{\left\langle e^{-ik \Delta r} \,
    \frac{1}{\tau} \left(-n - \sum_i \frac{\text{d}\ln U(y_i,x_i,\tau_i)}{\text{d}\tau_i} \, \tau_i \right)
    \right\rangle_\text{MC}}{\langle e^{-ik \Delta r} \rangle_\text{MC}} \, .
\end{equation}

\subsection{Effective mass estimator}

On one hand, taking the second log-derivative of (\ref{gf_asympt}) with respect to $k$ at $k=0$,
\begin{equation}\label{deriv1}
    \frac{1}{G(k,\tau)} \frac{\text{d}^2G(k,\tau)}{\text{d}k^2}_{k=0} =
    -\frac{\tau}{m^*}
    +\frac{\text{d}^2\ln(Z)}{\text{d}k^2}_{k=0} \, ,
\end{equation}
where we used the fact that the first derivative of $Z(k)$ vanishes at $k = 0$ because $Z(k)$ is an even function, and
\begin{equation}
    \frac{1}{m^*} = \frac{\text{d}^2E_0}{\text{d}k^2}_{k=0} \, .
\end{equation}
On the other hand one can transform the Green's function to position space
\begin{equation}\label{eq:mass-gf2}
    G(k,\tau) = \sum_{\Delta r} e^{-ik \Delta r} G(\Delta r, \tau) \, .
\end{equation}
Now we take the second log-derivative in $k$ of Eq.~(\ref{eq:mass-gf2}), and then substitute the expression (\ref{pert-exp})
\begin{align}\label{deriv2}
    \frac{1}{G(k,\tau)} \frac{\text{d}^2G(k,\tau)}{\text{d}k^2}_{k=0} =
    \frac{\sum_{\Delta r} -(\Delta r)^2 G(\Delta r, \tau)}{\sum_{\Delta r} G(\Delta r, \tau)} = \nonumber \\
    \frac{\sum_{\Delta r,\nu} -(\Delta r)^2 W_\nu(\Delta r, \tau)}{\sum_{\Delta r,\nu} W_\nu(\Delta r, \tau)} =
    \langle -(\Delta r)^2 \rangle_\text{MC} \, .
\end{align}
By equating Eq.~(\ref{deriv1}) and Eq.~(\ref{deriv2}) we arrive at the final expression for the effective mass estimator
\begin{equation}
    \frac{1}{m^*} -
    \frac{1}{\tau} \frac{\text{d}^2\ln(Z)}{\text{d}k^2}_{k=0}
    = \left\langle \frac{(\Delta r)^2}{\tau} \right\rangle_\text{MC} \, .
\end{equation}
The simulation is obviously restricted to a finite $\tau_{max}$, while we need to extrapolate to $\tau \to \infty$ to obtain $1/m^*$. Thus we perform a linear fit of the MC estimator versus $1/\tau$ and take the intercept as our unbiased value of the inverse effective mass.

%% file: Z_asymptotic.tex
\section{Analytical formula for asymptotic value of \texorpdfstring{$Z$}{Z}-factor at large \texorpdfstring{$|g_2|$}{|g2|}}
\label{assyan}

\begin{figure}[h]
    \centering
    \includegraphics[width=0.40\textwidth]{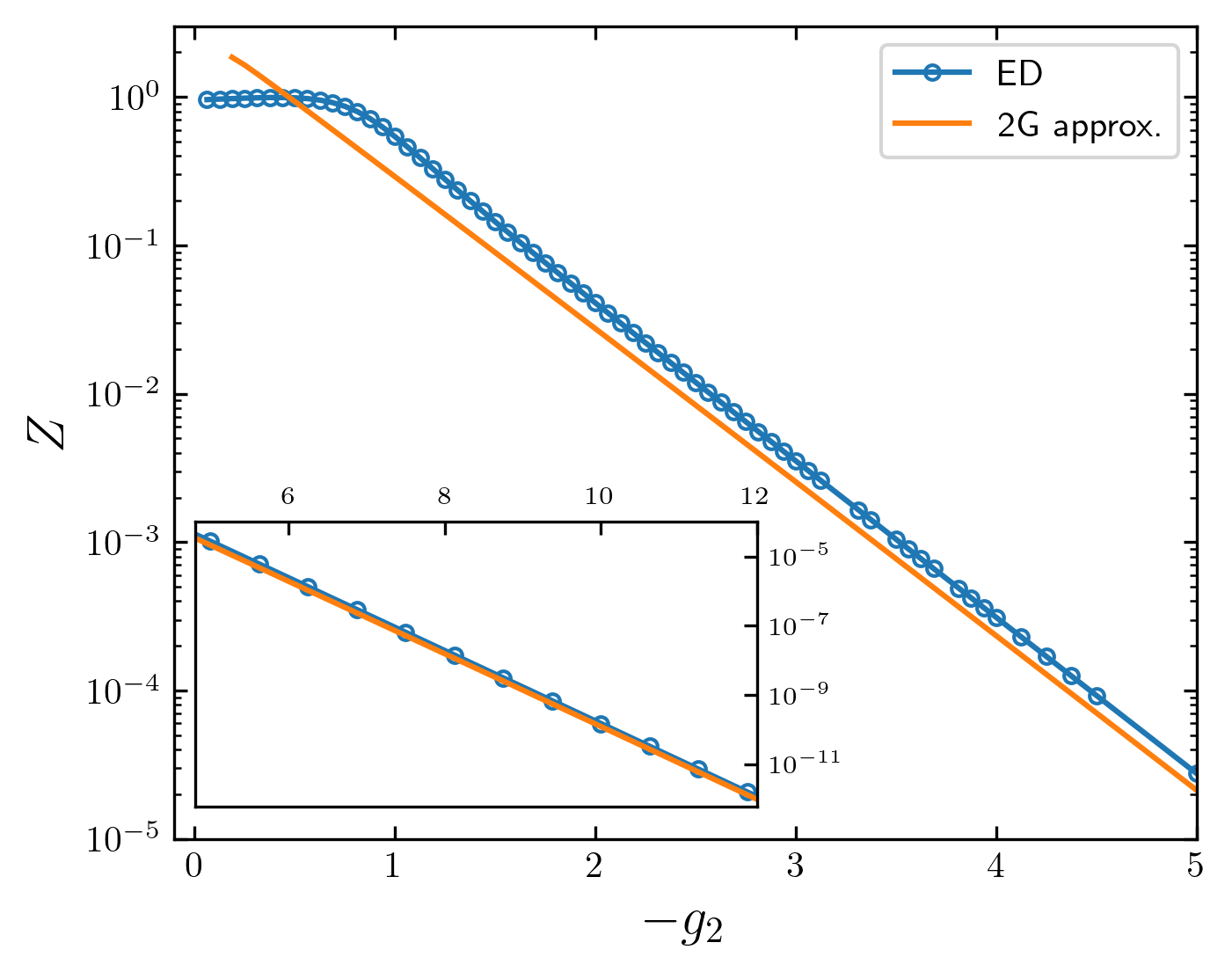}
    \caption{Comparison between analytical formula (\ref{Z_analytic}) labeled 2G (double-Gaussian) approximation, and the result from single-site ED, for $g_4 = 0.1$ and $\Omega = 0.5$. Even at $g_2=-12$, the error is 17\%.}
    \label{fig:el1}
\end{figure}

When $|g_2|$ is large, the two wells become deep and separated so that the oscillator's wave function can be approximated with a superposition of harmonic states centered in the double well's minima $\pm x_0$ (\ref{x00}) and with renormalized frequency $\tilde\omega$ ({\ref{tildeom}})
\begin{align}
    \psi_0(x) &= \left(\frac{\omega}{\pi}\right)^{1/4} \exp\left(-\frac{\omega x^2}{2}\right) \label{eq:HO} \, , \\ \label{eq:2G}
    \tilde \psi_0(x) &= \frac{\psi_{+} + \psi{-}}{\sqrt{2}} \,, \quad \\
    \tilde \psi_{\pm}(x) &= \left(\frac{\tilde \omega}{\pi}\right)^{1/4} \exp\left(-\frac{\tilde \omega (x \mp x_0)^2}{2}\right) \; .
\end{align}
Note that this calculation assumes a single-site system ($t=0$) or a system where $t \ll |g_2|$.

Computing the overlap 
\begin{align}
    \braket{\psi_0|\tilde \psi_{+}} &=
    \left( \frac{\omega\tilde\omega}{\pi^2} \right)^{1/4} \int \text{d}x \, 
    e^{-\omega x^2/2} \,
    e^{-\tilde \omega (x-x_0)^2/2} \\
    &= \left( \frac{R}{\pi^2} \right)^{1/4} \int \text{d}z \, 
    \exp\left(-\frac{z^2}{2} -\frac{R(z-z_0)^2}{2}\right) \\
    &= \left( \frac{R}{\pi^2} \right)^{1/4}
    \sqrt{\frac{2\pi}{1+R}}
    \exp\left(-\frac{R}{1+R} \, z_0^2\right) \, ,
\end{align}
one gets $Z$-factor $Z=\braket{\psi_0|\tilde \psi_0}^2$ which results in the final expression (\ref{Z_analytic}).

\begin{figure}[h]
    \centering
    \includegraphics[width=0.40\textwidth]{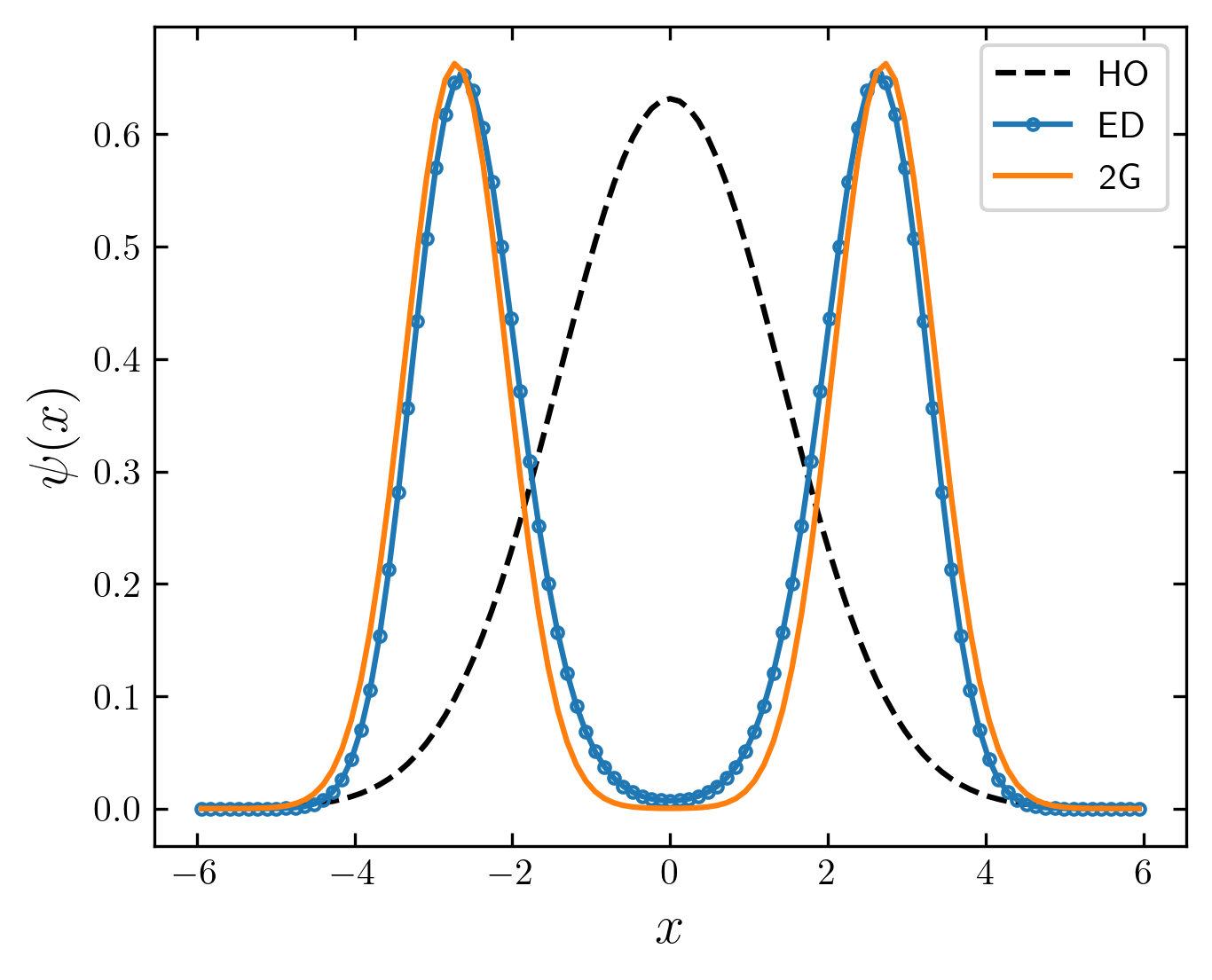}
    \caption{Comparison between the double-Gaussian (2G) approximation (\ref{eq:2G}) to the wavefunction at $g_2 = -1.6$ and the single-site ED result. The harmonic oscillator (HO) ground state (\ref{eq:HO}) is also shown for reference.}
    \label{fig:el2}
\end{figure}

Fig.~\ref{fig:el1} shows that the asymptotic formula is valid only at very large values of $|g_2|$. To better understand the reason for the discrepancy, in Fig.~\ref{fig:el2} we plot the wavefunctions (\ref{eq:HO}) and (\ref{eq:2G}) used to obtain the Z factor, at an intermediate value of the coupling $g_2 = -1.6$. Near $x=0$, where the ground state wavefunction of the harmonic oscillator is large, the double-Gaussian wavefunction is excessively suppressed, unlike the exact wavefunction obtained through single-site ED.

